\documentclass[%
 preprint,
 superscriptaddress,
 amsmath,amssymb,
 aps,
 showkeys,
 titlepage,
 endfloats*.   
]{revtex4-2}

\usepackage[utf8]{inputenc}
\usepackage[english]{babel}
\usepackage{booktabs}
\usepackage{makecell}

\usepackage{graphicx}
\usepackage{dcolumn}
\usepackage{bm}

\usepackage[colorlinks = true,
            linkcolor = blue,
            urlcolor  = blue,
            citecolor = red,
            anchorcolor = blue]{hyperref}

\DeclareUnicodeCharacter{0308}{\"{}}
\begin{document}


\title{High-Throughput Search for Photostrictive Materials based on a Thermodynamic Descriptor} 

\author{Zeyu Xiang}
\thanks{These authors contributed equally.}
\affiliation{Department of Mechanical Engineering, University of California, Santa Barbara, CA 93106, USA}

\author{Yubi Chen}
\thanks{These authors contributed equally.}
\affiliation{Department of Mechanical Engineering, University of California, Santa Barbara, CA 93106, USA}
\affiliation{Department of Physics, University of California, Santa Barbara, CA 93106, USA}

\author{Yujie Quan}
\affiliation{Department of Mechanical Engineering, University of California, Santa Barbara, CA 93106, USA}

\author{Bolin Liao}
\email{bliao@ucsb.edu} \affiliation{Department of Mechanical Engineering, University of California, Santa Barbara, CA 93106, USA}

\date{\today}

\begin{abstract}
Photostriction is a phenomenon that can potentially improve the precision of light-driven actuation, the sensitivity of photodetection, and the efficiency of optical energy harvesting. However, known materials with significant photostriction are limited, and effective guidelines to discover new photostrictive materials are lacking.
In this study, we perform a high-throughput computational search for new photostrictive materials based on simple thermodynamic descriptors, namely the band gap pressure and stress coefficients. Using constrained density functional theory simulations, we establish that these descriptors can accurately predict intrinsic photostriction in a wide range of materials.
Subsequently, we screen over 4770 stable semiconductors with a band gap below 2 eV from the Materials Project database to search for strongly photostrictive materials.
This search identifies PtS$_2$ and Te$_2$I as the most promising ones, with photostriction exceeding 10$^{-4}$ with a moderate photocarrier concentration of 10$^{18}$ cm$^{-3}$.
Furthermore, we provide a detailed analysis of factors contributing to strong photostriction, including bulk moduli and band-edge orbital interactions. 
Our results provide physical insights into photostriction of materials and demonstrate the effectiveness of using simple descriptors in high-throughput searches for new functional materials.
\end{abstract}

                            
\maketitle

\section{Introduction}

Photostriction, defined as the nonthermal mechanical strain of materials when exposed to light, has attracted significant research interests given important applications in photostrictive actuators~\cite{poosanaas2000photostrictive}, optical sensors~\cite{chen2021photostrictive}, information storage~\cite{liparo2023static}, and energy harvesting~\cite{lafont2012magnetostrictive}. Since its discovery in the 1960s, photostriction has been studied in a broad range of semiconductors. In nonpolar semiconductors, such as Si~\cite{gauster1967electronic} and Ge~\cite{figielski1961photostriction}, it is understood that photostriction originates from an ``electronic volume effect''(EVE)~\cite{north1966length}, where the occupation of valence and conduction bands alters chemical bonds and leads to strain. EVE should occur on the time scale of electron-phonon coupling ($\lesssim$ ps) and contributes to photostriction in all semiconducting materials. In polar materials without inversion symmetry, such as zinc-blende~\cite{lagowski1974photomechanical} and wurtzite~\cite{lagowski1972photomechanical} polar semiconductors and ferroelectric materials, the inverse piezoelectric effect (IPE) can provide an additional contribution to photostriction. For example, local screening of the polarization by photoexcited carriers in ferroelectrics can lead to ultrafast photostriction on a picosecond time scale due to IPE in ferroelectrics~\cite{schick2014localized,paillard2016photostriction}. In addition, macroscopic electron-hole separation induced by a surface photovoltage effect (SPV) in polar semiconductors~\cite{lagowski1974photomechanical} or a bulk photovoltage (BPV) effect in ferroelectrics can lead to strong photostriction through IPE, which, however, typically occurs on a much slower time scale set by the charge transport process~\cite{dogan2001nanocomposite}. In these inorganic semiconductors, strain induced by photostriction is typically below $10^{-5}$ at moderate photocarrier concentrations achievable under continuous laser illumination~\cite{kundys2015photostrictive}. Although amorphous chalcogenide glasses and organic semiconductors can exhibit giant photostrictive strain over 1\% due to photo-induced bond modification and photoisomerization~\cite{kuzukawa1998photoinduced,finkelmann2001new}, these effects are usually much slower, taking minutes to hours to finish. Here we make a distinction between ``intrinsic'' photostrictive effects caused by the presence of uniform photocarriers in the bulk of the sample, e.g. EVE and local polarization screening in ferroelectrics, and ``extrinsic'' photostrictive effects that rely on macroscopic photocarrier transport and the presence of sample surfaces, e.g. photostriction associated with SPV and BPV effects. Intrinsic photostrictive effects do not require photocarrier transport and, thus, are much faster than extrinsic photostrictive effects, but typically lead to a smaller strain.

In recent years, several materials have been discovered to show giant photostriction resulting from newly proposed mechanisms. For example, the next-generation solar cell material, methylammonium lead iodide (MAPbI$_3$), shows significant photostriction, where the light-induced lattice change is $>$ 0.12\,\% based on atomic force microscope measurements~\cite{zhou2016giant,lv2021giant}. 
This effect was attributed to the weakening of hydrogen bonding by photo-induced charge carriers.
 In addition, strontium ruthenate (SrRuO$_3$), a transition metal oxide with a perovskite crystal structure, was observed with a 1.12$\%$ photo-induced strain based on the Raman peak shifts~\cite{wei2017photostriction}. 
The strong light-induced deformation was attributed to the nonequilibrium of phonons in SrRuO$_3$. Although these experimental discoveries are intriguing and exciting, systematic theoretical guidelines to search for strongly photostrictive materials are still lacking, limiting the efficiency of developing new photostrictive applications.




On the theoretical modeling side, photostriction has recently been studied from first-principles by a constrained density functional theory (c-DFT) method~\cite{kaduk2012constrained}, through which a fixed electron concentration is excited from a chosen valence band state to a specific conduction band state~\cite{paillard2016photostriction} and the lattice is allowed to relax with the electron occupation as a constraint.
The resulting lattice change along selected crystal directions as a result of the photoexcited electron-hole pairs is evaluated as photostriction. This method was applied to study photostrictive coefficients in ferroelectric materials~\cite{paillard2016photostriction,paillard2017ab}, where it was shown to accurately capture the EVE and the polarization screening contributions to photostriction (the intrinsic photostriction), while the extrinsic photostriction caused by macroscopic charge separation cannot be modeled since a spatially uniform distribution of electron-hole pairs is assumed in c-DFT simulations. 
c-DFT has also been applied successfully to model photostriction in halide perovskites~\cite{peng2022tunable}.
These results suggest that c-DFT can be used as an effective benchmark calculation of intrinsic photostrictive effects for semiconductors, but its relatively high computational cost prohibits its use to broadly screen and discover new materials with a strong photostriction.



In this work, we conduct a high-throughput search for strongly photostrictive materials based on a thermodynamic descriptor, the band gap pressure or stress coefficient, which can be efficiently evaluated by ground-state DFT calculations. The band gap pressure coefficient was connected to the photostriction coefficient in early studies~\cite{figielski1961photostriction}. Here, we provide a derivation based on thermodynamic Maxwell relations and generalize the result to materials with anisotropic photostriction, where the band gap stress coefficient is more relevant. We then verify that these thermodynamic descriptors can be used to accurately predict the intrinsic photostrictive effects in a broad range of materials by comparing to c-DFT simulations~\cite{kaduk2012constrained}. 
Encouraged by this result, we compute the band gap pressure coefficients of 4770 semiconducting materials (band gap within 2 eV) using DFT to search for new photostrictive materials.
Among these materials, the band gap pressure coefficient of PtS$_2$ is found to be the highest (-49.17 meV/kB), whose photostrictive strain along the out-of-plane direction is nearly 18 times higher than that of MAPbI$_3$ with an identical photoexcited electron concentration. 
Upon closer examination of materials exhibiting substantial band gap pressure/stress coefficients, it becomes apparent that those with a longer bond length tend to display higher photostriction given the rapid decrease of their bulk moduli.
Furthermore, a simplified linear combination of atomic orbitals (LCAO) analysis coupled with the crystal orbital Hamilton population (COHP) method~\cite{dronskowski1993crystal,deringer2011crystal} is subsequently utilized to compare photostriction among materials exhibiting similar bulk moduli. 
It is found that materials with band edges formed by cations and anions with a small orbital energy difference tend to exhibit high deformation potentials that lead to a strong photostriction. Our work identifies several new strongly
photostrictive materials for potential applications and provides systematic guidelines for understanding photostriction in a wide range of semiconductors.

\section{Theory and Methods}


\subsection{Thermodynamics of Intrinsic Photostriction}

In the early literature, intrinsic photostriction was heuristically linked to the band gap pressure coefficient in isotropic materials~\cite{figielski1961photostriction}. Here, we provide a simple thermodynamic argument to derive this result and generalize it to anisotropic materials, where photo-induced strains along different crystallographic directions are distinct. We focus only on intrinsic photostriction caused by the presence of uniform photocarriers and electron-phonon coupling. Technically, accurate modeling of extrinsic photostriction requires a description of macroscopic photocarrier transport, which is highly complex and not suitable for a high-throughput study. Practically, intrinsic photostriction occurs on a much faster timescale (controlled by electron-phonon interaction) than extrinsic photostriction (limited by photocarrier transport) and is, thus, more suitable for many applications. In addition, since all semiconductors exhibit intrinsic photostriction, our result examines the lower limit of achievable photostriction, whereas extrinsic photostriction in certain groups of semiconductors can further increase the photo-induced strain.

First, the change of the Gibbs free energy of a unit cell with electrons photoexcited from the valence band maximum (VBM) to the conduction band minimum (CBM) can be expressed as:
\begin{equation}
    dG=Vdp-SdT+(\mu_e-\mu_h)dn,
    \label{pressure_eq}
\end{equation}
where $V$ is the unit cell volume, $p$ is the pressure, $S$ is the entropy, $T$ is the temperature, $\mu_e$ ($\mu_h$) is the quasi-Fermi level of photoexcited electrons (holes), and $n$ is the number of photoexcited electron-hole pairs per unit cell (assuming photoexcited electrons and holes have the same concentration). $\mu_e-\mu_h$ determines the open-circuit voltage generated by a semiconductor under light illumination~\cite{wurfel2016physics}.

According to the Maxwell's relation, the partial derivative of volume with respect to the excited electron number at a constant pressure ${\left(\frac{\partial V}{\partial n}\right)}_p$  is equal to the change of the open-circuit voltage with respect to pressure at a constant photocarrier concentration ${\left(\frac{\partial (\mu_e-\mu_h)}{\partial p}\right)}_n$:
\begin{equation}
    {\left(\frac{\partial V}{\partial n}\right)}_p={\left(\frac{\partial (\mu_e-\mu_h)}{\partial p}\right)}_n \label{1st_relation}.
\end{equation}
This relation suggests that how the open-circuit voltage evolves with increasing pressure at a constant photocarrier concentration can be used to predict the photostrictive behavior of materials. Since c-DFT examines the photostriction of materials at the ground state ($T=0$ K), where the open-circuit voltage approaches the band gap $E_G$, ${\left(\frac{\partial (\mu_e-\mu_h)}{\partial p}\right)}_n$ can be evaluated by the band gap pressure coefficient $\alpha_p={\left(\frac{\partial E_G}{\partial p}\right)}_n$. Namely, if we define a volumetric photostriction coefficient $V_n =  \left(\frac{\partial V}{\partial n}\right)_p$, it can be evaluated using the band gap pressure coefficient under a constant photocarrier concentration:
\begin{equation}
    \left(\frac{\partial V}{\partial n}\right)_p =  {\left(\frac{\partial E_G}{\partial p}\right)}_n \label{eqn:volume_form},  
\end{equation}
which can be efficiently computed using ground-state DFT.

More generally, photostriction can vary along different lattice directions in anisotropic materials, which leads us to rewrite the change of Gibbs free energy from Eq. \ref{pressure_eq} into the strain-stress form:
\begin{equation}
    dG=\sum_{ij}\epsilon_{ij}Vd\sigma_{ij}-SdT+E_Gdn, \label{stress_eq}
\end{equation}
where $\epsilon_{ij}$ and  $\sigma_{ij}$ refer to the strain and stress tensor components along three Cartesian directions, respectively. Here, the stress $\sigma_{ij}$ refers to the internal stress developed in the material as a response to an external load. 
Similarly, we define a linear photostriction coefficient as the partial derivative of a strain component along a specified lattice direction with respect to the  photoexcited electron concentration $\epsilon_n=V \left(\frac{\partial \epsilon_{ij} }{\partial n}\right)_{\sigma}$, which can then be related to the variation of the band gap with stress along the corresponding lattice direction $\alpha_s=\left(\frac{\partial E_G}{\partial \sigma_{ij}}\right)_{n,\,\mathrm{other}  \,\sigma_{ij}}$ under a constant photoexcited carrier concentration:
\begin{equation}
    V \left(\frac{\partial \epsilon_{ij} }{\partial n}\right)_{\sigma}=\left(\frac{\partial E_G}{\partial \sigma_{ij}}\right)_{n,\,\mathrm{other}  \,\sigma_{ij}},   \label{eqn:strain_form}
\end{equation}
where $\left(\frac{\partial \epsilon_{ij} }{\partial n}\right)_{\sigma}$ is evaluated with a constant stress along all directions (experimentally, stress-free condition along all directions), while $\left(\frac{\partial E_G}{\partial \sigma_{ij}}\right)_{n,\,\mathrm{other}  \,\sigma_{ij}}$ is evaluated when all other nonequivalent stress components are held constant. 
For brevity, we use $\left(\frac{\partial E_G}{\partial \sigma_{ij}}\right)_n$ in the remainder of this paper. In this work, we focus on photostrictive behaviors along the three main diagonal directions $(ij=xx,yy,zz)$, which are relevant for materials with tetragonal and orthorhombic symmetries.



\subsection{Computational Methods}

DFT calculations were accomplished via the Vienna ab initio simulation package (VASP)~\cite{kresse1996efficiency,kresse1996efficient} based on the projected augmented wave (PAW) pseudopotentials~\cite{blochl1994projector}. 
The Perdew-Burke-Ernzerhof form of generalized gradient approximation (PBE-GGA)~\cite{perdew1996generalized} was used for structural optimization. 
For materials studied in this paper, \textbf{k}-point convergence and plane-wave cut-off energy convergence were first tested to ensure that the lattice parameters were well relaxed. For example, for the calculation of MAPbI$_3$, the plane-wave cutoff energy was set at 520 eV and the energy and force convergence criteria were set to 1 $\times$ 10$^{-8}$ eV and 5 $\times$ 10$^{-4}$ eV/\AA, respectively. 
The $\Gamma$-centered $7\times 4\times 4$ \textbf{k}-point grids were used to sample the first Brillouin zone.

The c-DFT method~\cite{kaduk2012constrained} was used to simulate materials with a fixed concentration of photoexcited electrons, where the occupations of electronic states near the band edges were imposed as a constraint during the DFT simulation and lattice relaxation. 
To evaluate the first correlation ${\left(\frac{\partial V}{\partial n}\right)}_p={\left(\frac{\partial E_G}{\partial p}\right)}_n$, lattices were relaxed with different photocarrier concentrations under linearly distributed hydrostatic pressures, where the change of volume as a function of the photocarrier concentration was evaluated as the volumetric photostriction coefficient at a given pressure. On the right-hand side, the band gap was evaluated as a function of pressure at given photocarrier concentrations.
To evaluate the second correlation in anisotropic materials $V \left(\frac{\partial \epsilon_{ij} }{\partial n}\right)_{\sigma}={\left(\frac{\partial E_G}{\partial \sigma_{ij}}\right)}_n$, lattice relaxation was performed with different photocarrier concentrations to minimize stress along all directions. This constraint also corresponds to typical experimental conditions measuring photostriction of a stress-free sample. Lattice changes along different directions as a result of the relaxation were evaluated as the directional photostriction coefficients. On the right-hand side, the band gap was evaluated as a function of stress applied along one direction, while the lattice was relaxed to minimize stress along the other directions.


The bonding nature associated with electron energy bands was analyzed by the COHP method~\cite{dronskowski1993crystal}, which decomposes the energy of the electronic band structure into interactions (overlaps) between pairs of atomic orbitals between adjacent atoms. In other words, COHP analysis provides a bond-weighted electronic density of states and provides a quantitative measure of the bonding and anti-bonding contributions to the electronic bands, especially near the band extrema. In a COHP analysis, typically a positive (negative) sign corresponds to anti-bonding (bonding) interactions.
Conventionally, COHP diagrams plot the negative value (-pCOHP), therefore making bonding (anti-bonding) states on the right (left) of the axis for intuitive visualization~\cite{deringer2011crystal}.

For high-throughput search of new materials with significant photostriction, we selected 4770 stable materials with band gaps in the range of 0-2\,eV (calculated by DFT-PBE) from the Materials Project database~\cite{jain2013commentary}. 
A plane-wave cutoff energy of 400 eV was first utilized for the lattice relaxation of all materials to be screened, followed by band gap calculations under three different pressures [2 kilobar (kB), 6 kB, and 10 kB].
The extracted band gap pressure coefficients are then used to discover materials with potentially strong photostriction.
For promising material candidates, further \textbf{k}-point and plane-wave cutoff energy convergence were tested to ensure the convergence of the lattice parameters and forces, followed by COHP analysis of the bonding characteristics.
For example, the plane-wave cutoff energy was 500 eV and 600 eV for Te$_2$I and PtS$_2$, with energy and force convergence criteria set to 1 $\times$ 10$^{-8}$ eV and 1 $\times$ 10$^{-4}$ eV/\AA, and 1 $\times$ 10$^{-10}$ eV and 5 $\times$ 10$^{-6}$ eV/\AA., respectively.
Correspondingly, the $\Gamma$-centered $8\times 2\times 2$ and $11\times 11\times 6$ \textbf{k}-point grids were used to sample the first Brillouin zone of the two materials.


\section{Results and Discussions}

\subsection{Verification of Thermodynamic Descriptor for Isotropic and Anisotropic Materials}
\begin{figure}[h]
\includegraphics[width=
0.9\textwidth]{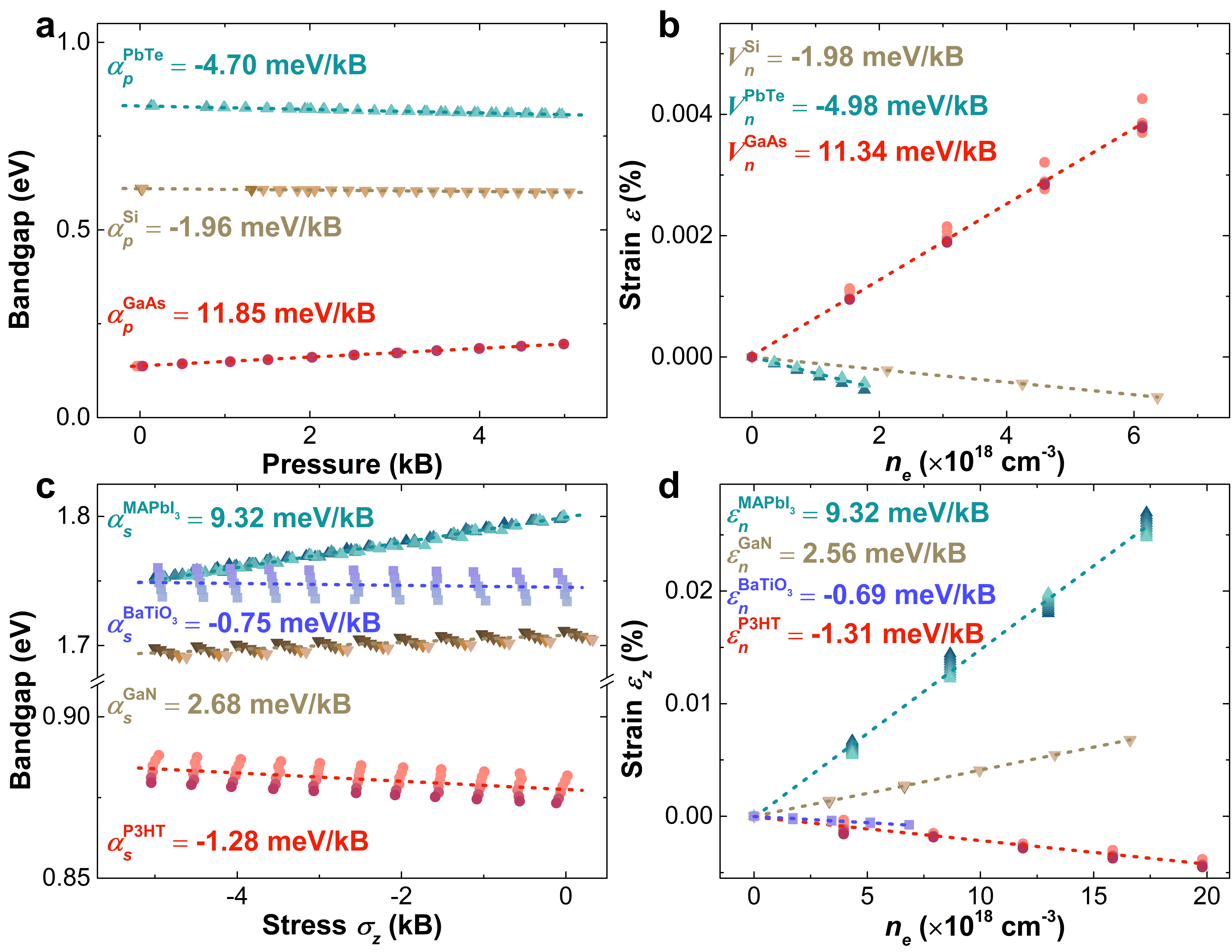}
\caption{\textbf{Relationship between the photostriction coefficients and the band gap pressure/stress coefficients.} \textbf{a}. Band gap changes with hydrostatic pressure at different excited electron densities corresponding to the range shown in \textbf{b} (represented by different shades of the marker color) for isotropic materials (Si, PbTe, and GaAs). 
Band gap pressure coefficient fittings (averaged over different excited electron densities) are denoted by dashed lines.
\textbf{b}. Photo-induced strain change with excited electron densities at different hydrostatic pressures corresponding to the range shown in \textbf{a} (represented by different shades of the marker color) for isotropic materials. 
Volumetric photostriction coefficient fittings (averaged at different hydrostatic pressures) are denoted by dashed lines.
\textbf{c}. Band gap changes with stress along the out-of-plane (z) direction with different excited electron densities for anisotropic materials (GaN, MAPbI$_3$, PT, and BaTiO$_3$). 
\textbf{d}. Photo-induced strain along the out-of-plane (z) direction as a function of excited electron densities for anisotropic materials.}
\label{fig:fig1}
\end{figure}

We first performed simulations to verify that the thermodynamic descriptors, i.e. the band gap pressure and stress coefficients in Eqn.~\ref{eqn:volume_form} and Eqn.~\ref{eqn:strain_form}, can indeed be used to accurately predict the photostrictive coefficients in isotropic and anisotropic materials evaluated by c-DFT. Several representative isotropic materials including Si (lattice parameter 5.47 \AA, non-polar), PbTe (lattice parameter 6.54 \AA, polar), and GaAs (lattice parameter 5.75 \AA, polar) were first studied. 
To stay in the linear regime where Maxwell's relation is valid, the number of excited electrons at a single \textbf{k}-point (at the VBM) per unit cell was limited to be less than 2 across materials (corresponding to a photocarrier concentration below 10$^{20}$ cm$^{-3}$), with the imposed hydrostatic pressure varying from 0 to 5 kB.
The difference of \textbf{k}-point weights between VBM and CBM was considered for Si given the indirect band gap.
The number of excited electrons at VBM per unit cell $n$ is transformed into an excited electron density $n_e$ after considering the \textbf{k}-point weight at VBM and the unit cell volume.
Figure~\ref{fig:fig1}a shows how the band gap changes with the hydrostatic pressure in the presence of different excited electron densities (marked by different shades of the symbols) for Si, PbTe, and GaAs.
Band gap pressure coefficients $\alpha_p={\left(\frac{\partial E_G}{\partial p}\right)}_n$ were fitted as -1.96 meV/kB, -4.70 meV/kB, and 11.85 meV/kB for Si, PbTe, and GaAs, respectively, and compared with previous experiments and calculations (-1.41 meV/kB~\cite{welber1975dependence}, -1.90 meV/kB~\cite{wei1999predicted} for Si; -7.40 meV/kB~\cite{dornhaus2006narrow} and -4.01 meV/kB~\cite{wei1997electronic} for PbTe; 12.60 meV/kB~\cite{welber1975dependence} and 9.8 meV/kB~\cite{wei1999predicted} for GaAs).
Notably, GaAs shows a positive band gap pressure coefficient, while Si and PbTe show negative band gap pressure coefficients. GaAs is a direct band gap semiconductor with strong covalent bonds, whose band gap is formed between a bonding valence band and an antibonding conduction band. In this case, an applied pressure reduces the interatomic distance and enhances the bonding-antibonding interaction, thus increasing the band gap~\cite{wei1999predicted}. In contrast, PbTe is a direct band gap semiconductor with an antibonding VBM~\cite{wei1997electronic,yuan2023lattice}, therefore reduced interatomic distance raises the VBM energy and reduces its band gap. 
On the other hand, Si is an indirect band gap semiconductor, whose band gap shrinks under pressure likely due to a broadened width of the conduction band~\cite{wei1999predicted}. 
Figure~\ref{fig:fig1}b presents the developed strain by excited electron densities at different hydrostatic pressures for the three materials, from which the values of volumetric photostriction coefficients $V_n={\left(\frac{\partial V}{\partial n}\right)}_{p}$ were determined as -1.98 meV/kB (Si), -4.98 meV/kB (PbTe), and 11.34 meV/kB (GaAs), respectively. A direct comparison of these results to experimentally measured photostriction is difficult due to the uncertainty in estimating the photocarrier concentration achieved in experiments. Assuming a moderate photocarrier concentration of 10$^{17\sim18}$ cm$^{-3}$, our predicted linear photostriction of selected materials is summarized in Table~\ref{tab:photostriction_comparison}. 
Here the linear photostriction $\frac{\Delta L}{L}$ for isotropic materials is determined by $\frac{1}{3}\frac{\Delta V}{V}$. For Si, a negative photostriction (photoinduced contraction) on the order of 10$^{-7}$ is in agreement with experiments~\cite{gauster1967electronic}. GaAs and CdS show positive photostriction (photo-induced expansion) with coefficients slightly higher than Si in our simulation. Photo-induced expansion in GaAs and CdS were experimentally observed~\cite{lagowski1972photomechanical,lagowski1974photomechanical}, but it is hard to compare the absolute magnitude due to extrinsic contributions from the SPV effect and thermal expansion, in addition to unknown experimental photocarrier concentrations. Importantly, however, the close agreement between $\alpha_p$ and $V_n$ in our simulations confirms that the band gap pressure coefficient can be used to accurately predict the intrinsic volumetric photostriction coefficient in isotropic materials evaluated by c-DFT.


\begin{table}[h]
    \centering
    \setlength{\tabcolsep}{5mm}
    \caption{Predicted photostriction of selected materials at a photocarrier concentration of $10^{17}$ cm$^{-3}$ to $10^{18}$ cm$^{-3}$.}
    \begin{tabular}{cc}
         \toprule
            & \thead{Simulated Photostriction, \\ $\Delta L/L$} \\
         \midrule
         Si & $-1.1\times10^{-7\sim-6}$  \\
         GaAs & $6.3\times10^{-7\sim-6}$  \\
         CdS & $2.1\times10^{-7\sim-6}$  \\
         BaTiO$_3$ (z-direction) & $-1.1\times10^{-7\sim-6}$ \\
         MAPbI$_3$ (x-direction) & $-1.6\times10^{-6\sim-5}$ \\
         MAPbI$_3$ (y-direction) & $-2.0\times10^{-6\sim-5}$ \\
         MAPbI$_3$ (z-direction) & $1.5\times10^{-6\sim-5}$ \\
         Te$_2$I (z-direction) & $-1.0\times10^{-5\sim-4}$ \\
         PtS$_2$ (z-direction) & $-2.7\times10^{-5\sim-4}$ \\
         \bottomrule
    \end{tabular}
    \label{tab:photostriction_comparison}
\end{table}

Anisotropic materials like GaN (wurtzite, $a$, $b$ = 3.22 \AA, $c$ = 5.24 \AA), MAPbI$_3$ (orthorhombic, $a$, $b$, $c$ = 8.56, 9.32, 12.92 \AA), polythiophene (PT) molecular crystal (orthorhombic, $a$, $b$, $c$ = 8.62, 6.00, 7.81 \AA)~\cite{cheng2019thermal} and BaTiO$_3$ (tetragonal, $a$, $b$,  = 3.99 \AA, $c$ = 4.10 \AA) were subsequently investigated to verify the thermodynamic relation given in Eqn.~\ref{eqn:strain_form}. 
The crystal structures and orientations of these materials are shown in Supplementary Material Section I.A. In these materials, photo-induced strain can vary significantly along different directions, therefore the volumetric photostriction coefficient alone is insufficient to fully describe the photo-induced strain state. 
The lattice structures were first relaxed under different hydrostatic pressures with the same step as isotropic materials to account for the photo-induced strain along different directions $\epsilon_n=V \left(\frac{\partial \epsilon_{ij} }{\partial n}\right)_{\sigma}$.
Then the band gap stress coefficient $\alpha_s=\left(\frac{\partial E_G}{\partial \sigma_{ij}}\right)_n$ was calculated by imposing strain along a specific direction (in the range of 0 to 0.02) and evaluating the corresponding stress along that direction while relaxing the stress along other directions with different excited electron numbers at VBM per unit cell (0 to 2).
Figure~\ref{fig:fig1}c shows the band gap change in these materials with stress along the out-of-plane ($\sigma_z$) direction with different excited electron densities for GaN, MAPbI$_3$, PT, and BaTiO$_3$ (results along the in-plane direction for GaN and MAPbI$_3$ are provided in Supplementary Material Section II.B. and II.C.).
The band gap stress coefficients $\alpha_s={\left(\frac{\partial E_G}{\partial \sigma_z}\right)}_{n}$ along the out-of-plane direction for GaN, MAPbI$_3$, PT, and BaTiO$_3$ were fitted as 2.68 meV/kB, 9.32 meV/kB, -1.28 meV/kB, and -0.75 meV/kB, respectively. 
Although it was previously reported that the band gap of MAPbI$_3$ was reduced with an increasing hydrostatic pressure~\cite{wang2015pressure}, the positive out-of-plane $\epsilon_n$ here is not contradicting the previous result due to the anisotropic response.
Figure\,\ref{fig:fig1}d depicts the corresponding strain change with excited electron densities, and the linear coefficients of photostriction $\epsilon_n={V\left(\frac{\partial \epsilon_{z}}{\partial n}\right)}_{\sigma}$ were calculated as 2.56 meV/kB (GaN), 9.32 meV/kB (MAPbI$_3$), -1.31 meV/kB (PT), and -0.69 meV/kB (BaTiO$_3$), accordingly. 
The results are summarized in Table~\ref{tab:photostriction_comparison}. 
For MAPbI$_3$ exposed to light, the out-of-plane direction expands and the photostriction evaluated here is $1.5\times10^{-6\sim-5}$ for a photocarrier concentration of 10$^{17}$ to 10$^{18}$ cm$^{-3}$, close to the experimental measurements of  $5\times10^{-5}$ in a bulk MAPbI$_3$ crystal~\cite{zhou2016giant}.
However, its in-plane lattice contracts under light with a photostriction on the order of $2\times10^{-6\sim-5}$.
In tetragonal BaTiO$_3$, the out-of-plane direction refers to the polarization direction, and 10$^{-4}$ compression of the polar axis under visible light was previously observed experimentally~\cite{bagri2022amalgamation}.
Here, we also predict a lattice contraction along the polarization direction of BaTiO$_3$, but the order of magnitude is hard to compare considering the uncertainties in estimating the photocarrier concentration.

\begin{figure}[!htb]
\includegraphics[width=0.7\textwidth]{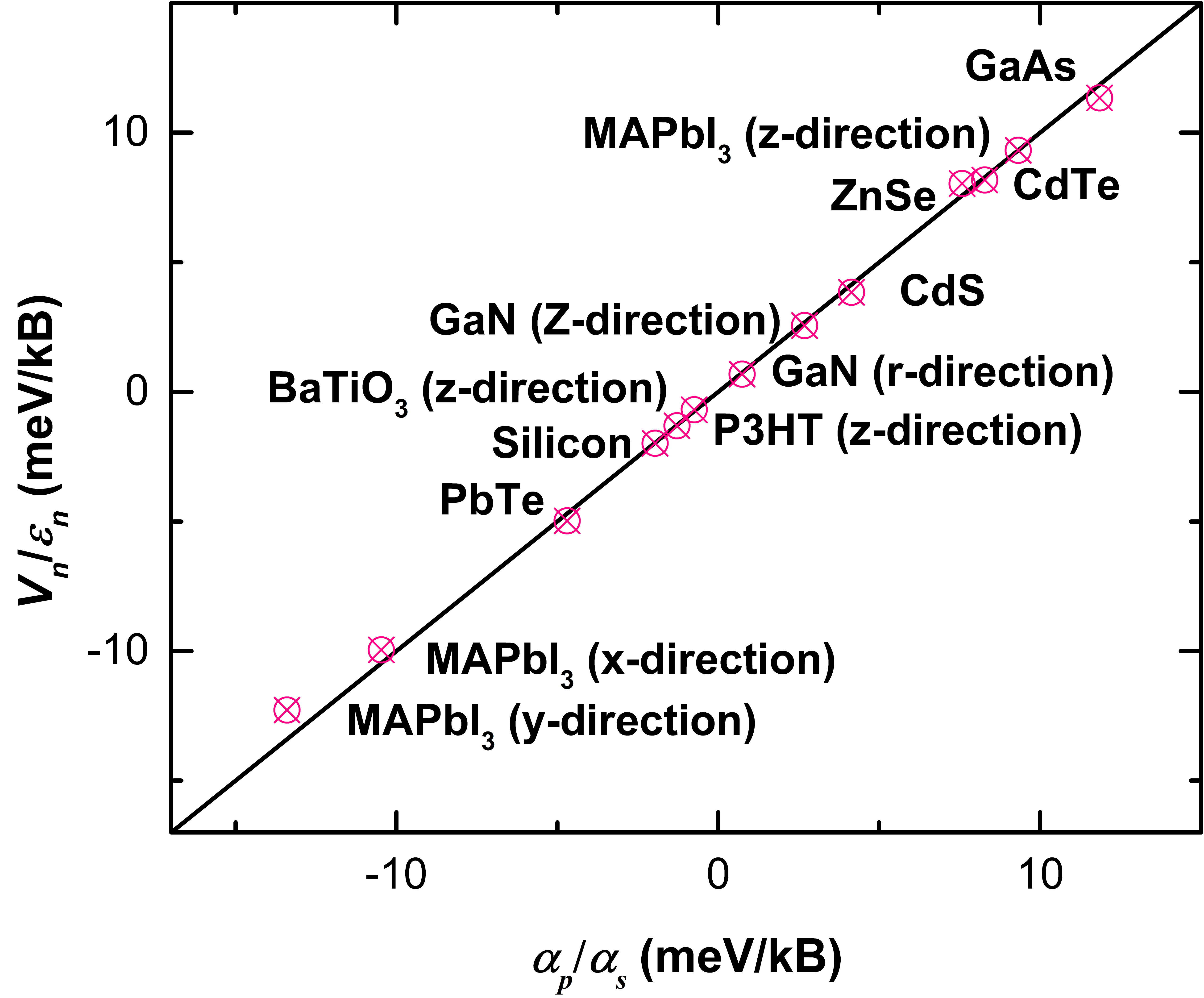}
\caption{\textbf{Comparison between the photostriction coefficients ($V_n/\epsilon_n$) and the band gap pressure (stress) coefficients ($\alpha_p/\alpha_s$) across different materials.}} 
\label{fig:fig2}
\end{figure}

Figure \ref{fig:fig2} summarizes the results comparing the photostriction coefficients simulated by c-DFT to the band gap pressure or stress coefficients evaluated by DFT for selected isotropic materials (Si, GaAs, CdTe, CdS, PbTe, ZnSe) and anisotropic materials (GaN, MAPbI$_3$, PT,  BaTiO$_3$). Detailed simulation results of these materials are provided in the Supplementary Material.
Among these materials, GaAs presents the highest band gap pressure coefficient of 11.85 meV/kB. 
ZnSe and CdTe possess band gap pressure coefficients of 7.58 meV/kB and 8.28 meV/kB, corresponding to a relatively large photostriction compared with CdS, whose band gap pressure coefficient is 4.14 meV/kB.
The slopes of band gap change with stress in $x$- and $y$-directions for orthorhombic MAPbI$_3$ are -10.47 and -13.40 meV/kB, indicating a lattice contraction under light illumination in these two directions.
The close agreement between the band gap pressure (stress) coefficient and photostriction coefficient, as presented in Figs.~\ref{fig:fig1} and ~\ref{fig:fig2}, suggests that the thermodynamic relation summarized in Eqns.~\ref{eqn:volume_form} and \ref{eqn:strain_form} is widely applicable to predict the intrinsic photostriction in a broad range of non-polar, polar, ferroelectric, and organic semiconductors.

\subsection{High-throughput Search of Materials with Promising Photostriction}
Following the validation of using the thermodynamic descriptors, i.e. band gap pressure and stress coefficients, to predict intrinsic photostriction coefficients, we conducted a high-throughput computational search of potential materials exhibiting significant photostriction thanks to the fact that the band gap pressure coefficients can be efficiently evaluated at the ground-state DFT level. 
We selected 4770 stable materials with band gaps ranging from 0 to 2 eV (as calculated by PBE DFT) from the Materials Project database~\cite{jain2013commentary}. 
Subsequently, the lattice structures were relaxed under a series of externally applied pressures (2 kB, 6 kB, and 10 kB), through which the band gap pressure coefficients were determined by linear fitting.
A screened list of the top 500 photostrictive materials is provided in Supplementary Material Section III.B.
Table~\ref{tab:top20} lists the band gap pressure coefficients and the relatively low bulk moduli (compared to the Si bulk modulus of 980 kB) of the top 20 materials, highlighting the crucial role of the latter in impacting the photostrictive behaviors of materials.
This aligns well with the physical intuition that materials with low bulk moduli are more easily distorted, which could be one indicator for strong photostriction.
In addition, we also analyzed the bonding nature of the VBM of these materials using the COHP method, which is also listed in Table~\ref{tab:top20}. Interestingly, we found that most materials with a large band gap pressure coefficient in our search exhibit an antibonding valence band. This observation suggests that an antibonding valence band may lead to a higher sensitivity of the band gap to external pressure or stress, and thus strong photostriction. Intuitively, occupied antibonding valence bands create instability in the lattice~\cite{yuan2023lattice}, and photoexcitation from antibonding valence bands may temporarily improve the stability of the lattice, leading to a large photostriction in the process. 
\begin{table}[]
    \centering
    \setlength{\tabcolsep}{5mm}
    \caption{Band gap pressure coefficients, bulk moduli, and VBM bonding type of the top 20 photostrictive materials discovered in our search.}
    \begin{tabular}{cccc}
         \toprule
          & \thead{Band gap Pressure Coefficient,\\ (meV/kB)}  & \thead{Bulk Modulus, \\ (kB)} &\thead{VBM, \\Bonding Type}\\
         \midrule
         PtS$_2$ & -49.17 & 71.84 & anti-bonding \\
         Te$_2$I & -48.53 & 95.14 & anti-bonding \\
         ScI$_3$ & -39.59 & 53.32 & anti-bonding \\
         HgI & -39.23 & 106.35 & anti-bonding \\
         Te$_2$Br & -38.71 & 106.45 & anti-bonding \\
         AsS & -38.30 & 60.35 & anti-bonding \\
         PdSe$_2$ & -35.78 & 148.48 & anti-bonding \\
         WS$_2$ & -33.71 & 104.23 & weak bonding \\
         PdS$_2$ & -33.54 & 123.37 & anti-bonding \\
         PAuS$_4$ & -33.23 & 51.91 & anti-bonding \\
         WSe$_2$ & -33.01 & 87.52 & weak bonding \\
         SiAs & -32.82 & 107.85 & anti-bonding \\
         Cs$_2$Se$_5$ & -32.26 & 77.60 & anti-bonding \\
         AuSe & -31.59 & 94.04 & anti-bonding \\
         PI$_2$ & -31.26 & 52.28 & anti-bonding \\
         InS & -31.14 & 165.11 & anti-bonding \\
         AsSe & -30.78 & 69.66 & anti-bonding \\
         MoSe$_2$ & -30.12 & 98.16 & weak bonding \\
         InSe & -29.78 & 109.33 & weak bonding \\
         TlBr$_2$ & -29.77 & 112.85 & anti-bonding \\
         
         \bottomrule
    \end{tabular}
    \label{tab:top20}
\end{table}

Among these materials of interest, PtS$_2$ and Te$_2$I present notably high band gap pressure coefficients of -49.17 meV/kB and -48.53 meV/kB, approximately 4 times that of MAPbI$_3$.
PtS$_2$ was previously synthesized and reported to have a CdI$_2$ structure, where each layer of the crystal is composed of a two-dimensional close-packed arrangement of metal atoms situated between two layers of chalcogen atoms~\cite{soled1976crystal,guo1986electronic,cullen2021synthesis}.
Te$_2$I belongs to tellurium sub-halides based on the substoichiometric halogen element, of which the crystalline structure is coordinated by threefold screw axis of Te element and shows interchain van der Waals (vdW) interactions~\cite{anastassakis1985raman}.
This material was first made through hydrothermal synthesis~\cite{kniep1976kenntnis}.
Considering their anisotropic and layered structures, we performed a detailed analysis of the out-of-plane photostriction for both materials.
Their crystal structures and directions are shown in Supplementary Material Section I.B.
Figure~\ref{fig:fig3}a and \ref{fig:fig3}c show that the band gap of Te$_2$I and PtS$_2$ decreases with stress $\sigma_{z}$ and the coefficients are -71.24 meV/kB and -173.46 meV/kB, respectively.
Figure~\ref{fig:fig3}b and \ref{fig:fig3}d present photo-induced lattice contraction of Te$_2$I and PtS$_2$ along the corresponding out-of-plane direction, where considerable photostriction coefficients are extracted as -63.73 meV/kB and -166.86 meV/kB (approximately 7 times and 18 times higher than MAPbI$_3$'s 9.32 meV/kB).
The small difference between the band gap stress coefficient and the photostriction coefficient for both Te$_2$I and PtS$_2$ once again demonstrates the robustness of the thermodynamic relations (Eqns.~\ref{eqn:volume_form} and \ref{eqn:strain_form}) in predicting photostriction in materials. 
Despite similar band gap pressure coefficients, the lattice of PtS$_2$ contracts significantly more along the out-of-plane direction compared to that of Te$_2$I under light illumination. Coefficients of photostriction of PtS$_2$ and Te$_2$I along the out-of-plane direction with a photocarrier concentration of 10$^{17}$ cm$^{-3}$ to 10$^{18}$ cm$^{-3}$ are $-2.7\times10^{-5\sim-4}$ and $-1.0\times10^{-5\sim-4}$, respectively, and compared to other materials in Table~\ref{tab:photostriction_comparison}. These photostriction coefficients are an order of magnitude higher than MAPbI$_3$ and two orders of magnitude higher than Si.
The remarkably large photostriction of PtS$_2$ and Te$_2$I defies the conventional impression that the intrinsic photostriction of inorganic bulk semiconductors is below $10^{-5}$.
The extraordinary photostriction of PtS$_2$ and Te$_2$I suggests their potential for advanced optomechanical applications.

\begin{figure}[!htb]
\includegraphics[width=\textwidth]{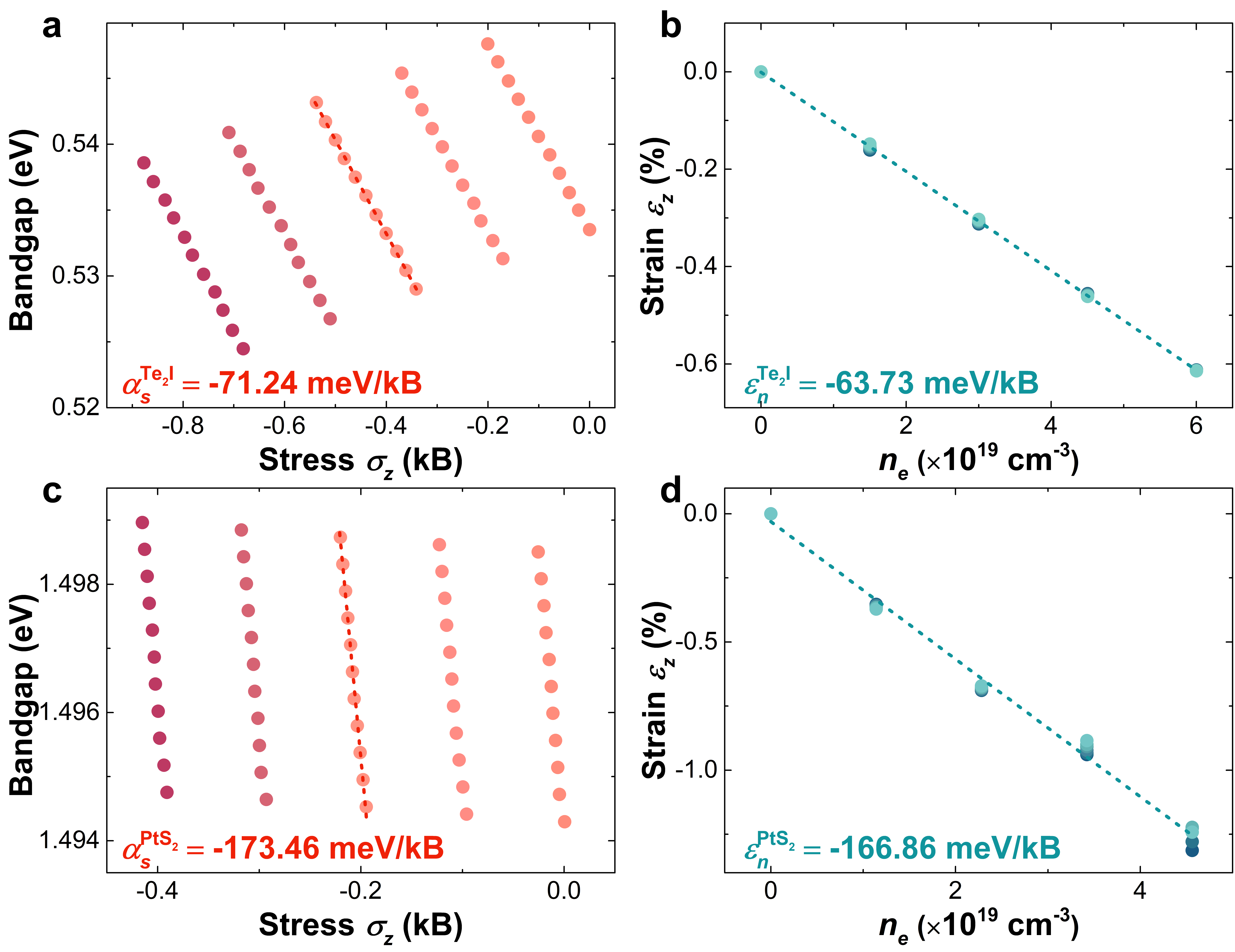}
\caption{\textbf{Photostriction and band gap stress coefficients in Te$_2$I and PtS$_2$.} \textbf{a}. Te$_2$I band gap changes with stress along the out-of-plane (z) direction with different excited electron densities corresponding to the range shown in \textbf{b} (represented by different shades of the marker color). The band gap stress coefficient averaged over different excited electron densities (dashed line) is fitted as -71.24 meV/kB. \textbf{b}. Te$_2$I strain change along the out-of-plane direction as a function of excited electron densities at different stress corresponding to the range in \textbf{a} (represented by different shades of the marker color). The photostriction coefficient averaged over different stress (dashed line) is fitted as -63.73 meV/kB. Similar plots for PtS$_2$ are presented in \textbf{c} and \textbf{d}.} 
\label{fig:fig3}
\end{figure}




\subsection{Physical Insights into Giant Photostriction}
\begin{figure}[!htb]
\includegraphics[width=\textwidth]{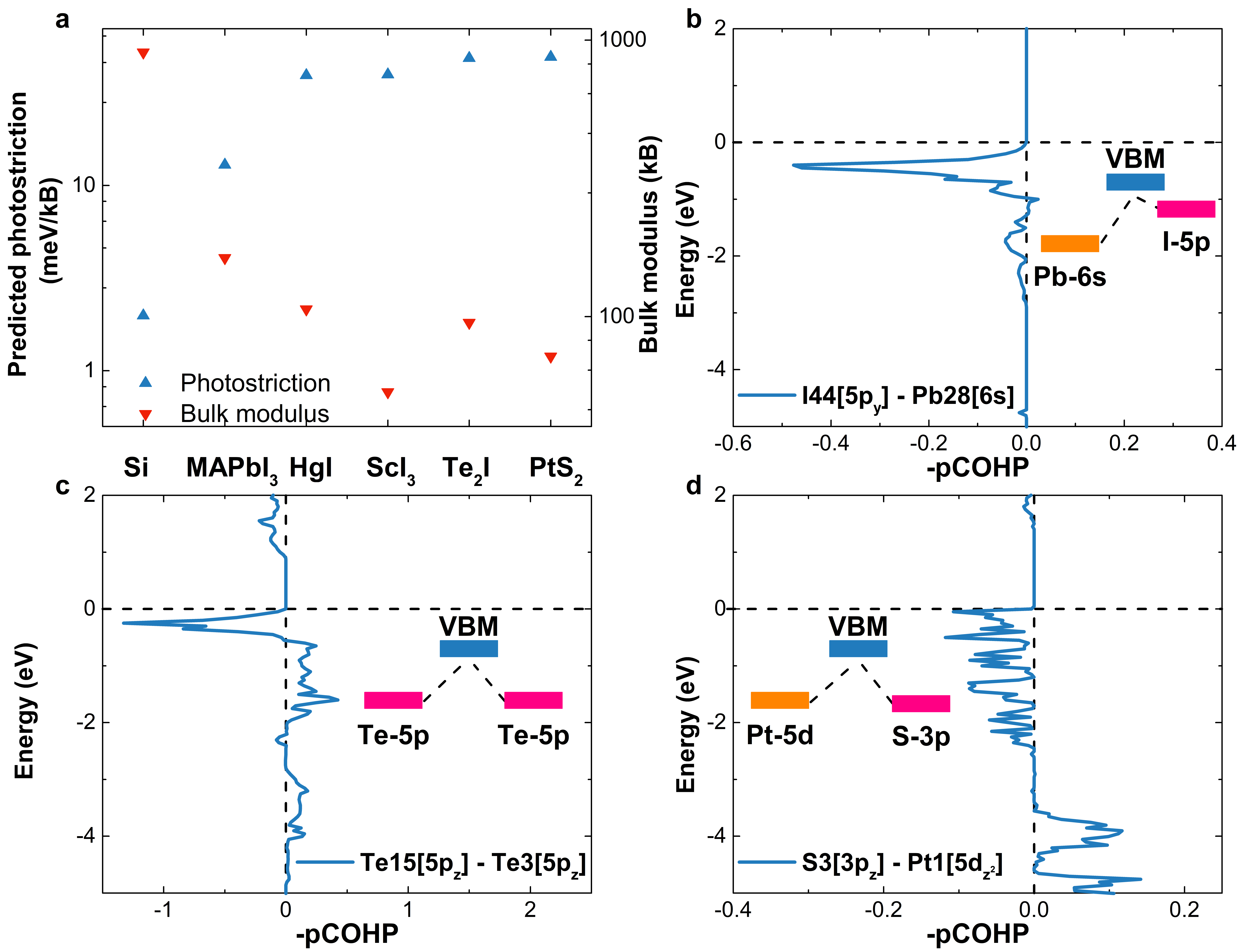}
\caption{\textbf{Analysis of factors impacting photostriction}. \textbf{a}. Predicted photostriction and the bulk modulus of Si, MAPbI$_3$, HgI, ScI$_3$, Te$_2$I, and PtS$_2$. \textbf{b-d}. COHP diagrams of MAPbI$_3$, Te$_2$I, and PtS$_2$, with insets highlighting strongest anti-bonding interactions between orbitals near VBM.} 
\label{fig:fig4}
\end{figure}

According to the thermodynamic relation (Eqn.~\ref{eqn:volume_form} and \ref{eqn:strain_form}), materials exhibiting giant intrinsic photostriction should also possess large band gap pressure coefficients $\alpha_p$, which can be expressed as~\cite{wei1999predicted}: 
\begin{equation}
    \alpha_p=\left(\frac{dE_G}{dp}\right)_n=-\frac{1}{B}\left(\frac{dE_G}{dlnV}\right)_n=-\frac{1}{B}\alpha_v \label{alpha_p}
\end{equation}
where $B=-\frac{dp}{dlnV}$ is the bulk modulus and $\alpha_v=\frac{dE_G}{dlnV}=\frac{dE_C}{dlnV}-\frac{dE_V}{dlnV}$ is the band gap volume coefficient, which is determined by the difference between the deformation potentials associated with CBM ($\frac{dE_C}{dlnV}$) and VBM ($\frac{dE_V}{dlnV}$) ($E_C$ and $E_V$ are the electronic energies of the CBM and VBM, respectively). A list of the band gap pressure and volume coefficients of semiconductors studied here is provided in Table~\ref{tab:table3}.
Equation~\ref{alpha_p} illustrates that a small bulk modulus, coupled with a significant difference in the deformation potentials between CBM and VBM, can lead to a substantial band gap pressure coefficient and, thus, a strong photostriction.
An analysis of materials included in our high-throughput screening reveals that the band gap pressure coefficient exhibits a generally decreasing trend with an increasing bulk modulus, which in turn decreases with an increasing bond length (see Fig. S9 in the Supplementary Material).
However, bulk modulus is not the sole factor - the band gap pressure coefficients of materials with similar bond lengths can still vary by two orders of magnitude due to the difference in their deformation potentials.
Wei and Zunger~\cite{wei1999predicted} adopted a simplified LCAO model to analyze the deformation potential associated with the band edges:
\begin{equation}
    E_{V/C}=\frac{\mu^c+\mu^a}{2}\mp\sqrt{(\frac{\mu^c-\mu^a}{2})^2+M^2} \label{orbital energy}
\end{equation}
where $\mu^c$ and $\mu^a$ are isolated cation and anion orbital energies before hybridization, respectively,
$-$ and $+$ represent bonding and anti-bonding interactions,
and $M$ is the matrix element between orbitals and inversely proportional to the bond length $l$ as $M \sim l^{-2}$~\cite{harrison2012electronic}.
Taking a derivative of the volume $V \sim l^3$, the deformation potential for VBM and CBM based on Eqn.~\ref{orbital energy} is given by:
\begin{equation}
    \alpha_v^{VBM/CBM}\sim\mp\frac{M^2}{\sqrt{(\frac{\mu^c-\mu^a}{2})^2+M^2}},
\end{equation}
which implies the magnitude of the deformation potential becomes larger when the orbital energy difference $\mu^c-\mu^a$ decreases.
Figure~\ref{fig:fig4}a compares the predicted magnitude of photostriction based on the band gap pressure coefficient (left y-axis) and the bulk moduli (right y-axis) among Si, MAPbI$_3$, HgI, ScI$_3$, Te$_2$I, and PtS$_2$, indicating that a relatively low bulk modulus is crucial for a significant photostriction. 
This is consistent with our physical intuition, as materials with lower bulk moduli are more prone to distortion under external loads. 
For MAPbI$_3$, Te$_2$I, and PtS$_2$ with comparable bulk moduli, the deformation potentials become essential when comparing the photostriction coefficient.
Figure~\ref{fig:fig4}b-d show the strongest anti-bonding interactions near VBM for these three materials based on COHP analyses.
The orbital energy differences for Te$_2$I ($\mu^{Te_{5p}}-\mu^{Te_{5p}}$) and PtS$_2$ ($\mu^{Pt_{5d}}-\mu^{S_{3p}}$) are relatively smaller compared to MAPbI$_3$ ($\mu^{I_{5p}}-\mu^{Pb_{6s}}$), leading to remarkably larger photostriction coefficients.

Based on the analysis presented above, we summarize our findings according to Table~\ref{tab:table3}, some of which follow those by Wei and Zunger~\cite{wei1999predicted}: (1) the magnitude of the band gap pressure coefficient $|\alpha_p|$ increases with periodically increasing anion atomic number due to the rapid decrease in the bulk modulus $B$. 
For example, $\alpha_p=-0.34$ meV/kB for AgCl and -0.75 meV/kB for AgBr; 4.14 meV/kB for CdS and 8.90 meV/kB for CdTe; 3.97 meV/kB for GaN and 11.67 meV/kB for GaAs; -12.18 meV/kB for PdBr$_2$ and -17.10 meV/kB for PdI$_2$; 5.94 meV/kB for ZnSe and 9.93 meV/kB for ZnTe.
The decrease in the orbital energy difference $\mu^c-\mu^a$, as the anion atomic number increases, is partially canceled by an increasing bond length $l$ contributing to the deformation potential $\alpha_v$.
(2) $|\alpha_p|$ decreases with periodically increasing cation atomic number due to the decrease in the deformation potential over the bulk modulus, which is attributed to increasing $\mu^c-\mu^a$ and $l$. 
For example, $\alpha_p=7.58$ meV/kB for ZnSe, 4.45 meV/kB for CdSe, 1.05 meV/kB for FeSe$_2$, 0.39 meV/kB for RuSe$_2$, -16.73 meV/kB for GeSe, -6.04 meV/kB for PbSe; -28.46 meV/kB for SiAs$_2$ and -20.53 for GeAs$_2$.
(3) The first two conclusions fail when the optical transitions change from direct $\Gamma-\Gamma$ to indirect processes due to the typically stronger matrix elements $M$ at $\Gamma$ point compared with other $\mathbf{k}$ points.
For instance, $\Gamma-L$ band gap pressure coefficient of GaP is 3.80 meV/kB, which falls between GaN (3.97 meV/kB) and GaAs (11.67 meV/kB) with $\Gamma-\Gamma$ transitions. 
Additionally, it is lower than $\alpha_p=8.46$ meV/kB for InP with a $\Gamma-\Gamma$ direct band gap.
(4) Ionic crystals exhibit smaller $|\alpha_v|$ in contrast to covalent ones of similar bond length because of the larger $\mu^c-\mu^a$ and smaller $M$.
For example, $\alpha_v=-2.99$ eV and 0.13 eV for CdTe and BaTe$_3$ with a similar bond length (3.30 \AA), $\alpha_v=0.54$ eV and 3.23 eV for Na$_2$AgSb and SiAs$_2$ with a similar bond length (2.99 \AA~and 2.98 \AA).
(5) Molecules with multiple identical cations or anions display larger $|\alpha_v|$ than those with a single cation or anion, given that the band edge is typically formed by hybridization among identical atomic species.
For example, $\alpha_v=4.56$ eV and 0.45 eV for Te$_2$I and TeI, $\alpha_v=2.46$ eV and 2.83 eV for GeAs and GeAs$_2$, $\alpha_v=0.16$ eV and 2.19 eV for CaP and CaP$_3$. These intuitive rules can serve as guidelines to rationally discover and design materials with an enhanced band gap pressure coefficient and photostriction.
  
\begin{table}[]
    \vspace{-10pt}
    \centering
    \setlength{\tabcolsep}{5mm}
    \caption{The band gap pressure coefficient and the band gap volume coefficient of selected semiconductors.}
    \begin{tabular}{ccc}
         \toprule
          & \thead{Band Gap Pressure Coefficient,\\ (meV/kB)}  & \thead{Band Gap Volume Coefficient, \\ (eV)} \\
         \midrule
         AgCl & -0.34 & 0.16 \\
         AgBr & -0.75 & 0.33 \\
         BaTe$_3$ & -0.51 & 0.13 \\
         CaP & -0.29 & 0.16  \\
         CaP$_3$ & -4.17 & 0.16  \\
         CdS & 4.14 & -2.24 \\
         CdSe & 4.45 & -2.23 \\
         CdTe & 8.90 & -2.99 \\
         FeSe$_2$ & 1.05 & -1.19 \\
         GaN & 3.97 & -6.75  \\
         GaP & 3.80 & -3.03 \\
         GaAs & 11.67 & -7.30 \\
         GeAs & -16.75 & 2.46 \\
         GeAs$_2$ & -20.53 & 2.83 \\
         GeSe & -16.73 & 3.87 \\
         InP & 8.46 & -5.45  \\
         Na$_2$AgSb & -1.89 & 0.54 \\
         PdBr$_2$ & -12.18 & 0.83 \\
         PdI$_2$ & -17.10 & 1.63 \\
         PbSe & -6.04 & 3.03 \\
         RuSe$_2$ & 0.39 & -0.56 \\
         SiAs$_2$ & -28.46 & 3.23 \\
         TeI & -5.83 & 0.45 \\
         Te$_2$I & -48.53 & 4.56 \\
         ZnSe & 5.94 & -3.40 \\
         ZnTe & 9.93 & -4.21 \\
         \bottomrule
    \end{tabular}
    \label{tab:table3}
\end{table}

\section{Conclusion}

In summary, through first-principles DFT simulations, we showed that simple thermodynamic descriptors, i.e. the band gap pressure and stress coefficients, can be used to accurately predict the intrinsic photostrictive coefficients in isotropic and anisotropic semiconductors. Motivated by this finding, we conducted a high-throughput search for promising photostrictive materials by screening over 4770 semiconducting materials with band gaps ranging from 0 to 2 eV, where we identified strong photostriction in PtS$_2$ and Te$_2$I that is at least one order of magnitude higher than that in MAPbI$_3$.
Furthermore, we analyzed in detail the impact of bulk modulus and deformation potentials on determining the band gap pressure coefficient and the photostriction.
We found that relatively low bulk moduli, which decrease with increasing bond lengths, are an important factor, agreeing with the physical intuition that materials with lower bulk moduli are more susceptible to distortion under external influences. 
For materials with similar bulk moduli, a simplified LCAO model was adopted for the analysis of the deformation potentials. 
We concluded that small orbital energy differences $\mu^c-\mu^a$ and large matrix elements $M$ at the band edges are favorable for achieving a large band gap pressure coefficient and a strong photostriction.
Our work provides guidelines for understanding intrinsic photostrictive behaviors from the point of view of band gap changes with external pressure and stress and predicts new promising photostrictive materials for optomechanical applications.

\begin{acknowledgments}
This work is based on research supported by the U.S. Office of Naval Research under the award number N00014-22-1-2262. Y.C. also acknowledges the support from the Graduate Traineeship Program of the NSF Quantum Foundry via the Q-AMASE-i program under award number DMR-1906325 at the University of California, Santa Barbara (UCSB). 
This work used Stampede2 at Texas Advanced Computing Center (TACC) and Expanse at San Diego Supercomputer Center (SDSC) through allocation MAT200011 from the Advanced Cyberinfrastructure Coordination Ecosystem: Services \& Support (ACCESS) program, which is supported by National Science Foundation grants 2138259, 2138286, 2138307, 2137603, and 2138296. Use was also made of computational facilities purchased with funds from the National Science Foundation (CNS-1725797) and administered by the Center for Scientific Computing (CSC). The CSC is supported by the California NanoSystems Institute and the Materials Research Science and Engineering Center (MRSEC; NSF DMR 2308708) at UCSB. 
\end{acknowledgments}

\bibliography{references.bib}

\begin{thebibliography}{43}%
\makeatletter
\providecommand \@ifxundefined [1]{%
 \@ifx{#1\undefined}
}%
\providecommand \@ifnum [1]{%
 \ifnum #1\expandafter \@firstoftwo
 \else \expandafter \@secondoftwo
 \fi
}%
\providecommand \@ifx [1]{%
 \ifx #1\expandafter \@firstoftwo
 \else \expandafter \@secondoftwo
 \fi
}%
\providecommand \natexlab [1]{#1}%
\providecommand \enquote  [1]{``#1''}%
\providecommand \bibnamefont  [1]{#1}%
\providecommand \bibfnamefont [1]{#1}%
\providecommand \citenamefont [1]{#1}%
\providecommand \href@noop [0]{\@secondoftwo}%
\providecommand \href [0]{\begingroup \@sanitize@url \@href}%
\providecommand \@href[1]{\@@startlink{#1}\@@href}%
\providecommand \@@href[1]{\endgroup#1\@@endlink}%
\providecommand \@sanitize@url [0]{\catcode `\\12\catcode `\$12\catcode `\&12\catcode `\#12\catcode `\^12\catcode `\_12\catcode `\%12\relax}%
\providecommand \@@startlink[1]{}%
\providecommand \@@endlink[0]{}%
\providecommand \url  [0]{\begingroup\@sanitize@url \@url }%
\providecommand \@url [1]{\endgroup\@href {#1}{\urlprefix }}%
\providecommand \urlprefix  [0]{URL }%
\providecommand \Eprint [0]{\href }%
\providecommand \doibase [0]{https://doi.org/}%
\providecommand \selectlanguage [0]{\@gobble}%
\providecommand \bibinfo  [0]{\@secondoftwo}%
\providecommand \bibfield  [0]{\@secondoftwo}%
\providecommand \translation [1]{[#1]}%
\providecommand \BibitemOpen [0]{}%
\providecommand \bibitemStop [0]{}%
\providecommand \bibitemNoStop [0]{.\EOS\space}%
\providecommand \EOS [0]{\spacefactor3000\relax}%
\providecommand \BibitemShut  [1]{\csname bibitem#1\endcsname}%
\let\auto@bib@innerbib\@empty
\bibitem [{\citenamefont {Poosanaas}\ \emph {et~al.}(2000)\citenamefont {Poosanaas}, \citenamefont {Tonooka},\ and\ \citenamefont {Uchino}}]{poosanaas2000photostrictive}%
  \BibitemOpen
  \bibfield  {author} {\bibinfo {author} {\bibfnamefont {P.}~\bibnamefont {Poosanaas}}, \bibinfo {author} {\bibfnamefont {K.}~\bibnamefont {Tonooka}},\ and\ \bibinfo {author} {\bibfnamefont {K.}~\bibnamefont {Uchino}},\ }\bibfield  {title} {\bibinfo {title} {Photostrictive actuators},\ }\href@noop {} {\bibfield  {journal} {\bibinfo  {journal} {Mechatronics}\ }\textbf {\bibinfo {volume} {10}},\ \bibinfo {pages} {467} (\bibinfo {year} {2000})}\BibitemShut {NoStop}%
\bibitem [{\citenamefont {Chen}\ and\ \citenamefont {Yi}(2021)}]{chen2021photostrictive}%
  \BibitemOpen
  \bibfield  {author} {\bibinfo {author} {\bibfnamefont {C.}~\bibnamefont {Chen}}\ and\ \bibinfo {author} {\bibfnamefont {Z.}~\bibnamefont {Yi}},\ }\bibfield  {title} {\bibinfo {title} {Photostrictive effect: characterization techniques, materials, and applications},\ }\href@noop {} {\bibfield  {journal} {\bibinfo  {journal} {Advanced Functional Materials}\ }\textbf {\bibinfo {volume} {31}},\ \bibinfo {pages} {2010706} (\bibinfo {year} {2021})}\BibitemShut {NoStop}%
\bibitem [{\citenamefont {Liparo}\ \emph {et~al.}(2023)\citenamefont {Liparo}, \citenamefont {Jay}, \citenamefont {Dubreuil}, \citenamefont {Simon}, \citenamefont {Fessant}, \citenamefont {Jahjah}, \citenamefont {Le~Grand}, \citenamefont {Sheppard}, \citenamefont {Prinsloo}, \citenamefont {Vlaminck} \emph {et~al.}}]{liparo2023static}%
  \BibitemOpen
  \bibfield  {author} {\bibinfo {author} {\bibfnamefont {M.}~\bibnamefont {Liparo}}, \bibinfo {author} {\bibfnamefont {J.-P.}\ \bibnamefont {Jay}}, \bibinfo {author} {\bibfnamefont {M.}~\bibnamefont {Dubreuil}}, \bibinfo {author} {\bibfnamefont {G.}~\bibnamefont {Simon}}, \bibinfo {author} {\bibfnamefont {A.}~\bibnamefont {Fessant}}, \bibinfo {author} {\bibfnamefont {W.}~\bibnamefont {Jahjah}}, \bibinfo {author} {\bibfnamefont {Y.}~\bibnamefont {Le~Grand}}, \bibinfo {author} {\bibfnamefont {C.}~\bibnamefont {Sheppard}}, \bibinfo {author} {\bibfnamefont {A.~R.}\ \bibnamefont {Prinsloo}}, \bibinfo {author} {\bibfnamefont {V.}~\bibnamefont {Vlaminck}}, \emph {et~al.},\ }\bibfield  {title} {\bibinfo {title} {Static and dynamic magnetization control of extrinsic multiferroics by the converse magneto-photostrictive effect},\ }\href@noop {} {\bibfield  {journal} {\bibinfo  {journal} {Communications Physics}\ }\textbf {\bibinfo {volume} {6}},\ \bibinfo {pages} {356} (\bibinfo {year} {2023})}\BibitemShut {NoStop}%
\bibitem [{\citenamefont {Lafont}\ \emph {et~al.}(2012)\citenamefont {Lafont}, \citenamefont {Gimeno}, \citenamefont {Delamare}, \citenamefont {Lebedev}, \citenamefont {Zakharov}, \citenamefont {Viala}, \citenamefont {Cugat}, \citenamefont {Galopin}, \citenamefont {Garbuio},\ and\ \citenamefont {Geoffroy}}]{lafont2012magnetostrictive}%
  \BibitemOpen
  \bibfield  {author} {\bibinfo {author} {\bibfnamefont {T.}~\bibnamefont {Lafont}}, \bibinfo {author} {\bibfnamefont {L.}~\bibnamefont {Gimeno}}, \bibinfo {author} {\bibfnamefont {J.}~\bibnamefont {Delamare}}, \bibinfo {author} {\bibfnamefont {G.}~\bibnamefont {Lebedev}}, \bibinfo {author} {\bibfnamefont {D.}~\bibnamefont {Zakharov}}, \bibinfo {author} {\bibfnamefont {B.}~\bibnamefont {Viala}}, \bibinfo {author} {\bibfnamefont {O.}~\bibnamefont {Cugat}}, \bibinfo {author} {\bibfnamefont {N.}~\bibnamefont {Galopin}}, \bibinfo {author} {\bibfnamefont {L.}~\bibnamefont {Garbuio}},\ and\ \bibinfo {author} {\bibfnamefont {O.}~\bibnamefont {Geoffroy}},\ }\bibfield  {title} {\bibinfo {title} {Magnetostrictive--piezoelectric composite structures for energy harvesting},\ }\href@noop {} {\bibfield  {journal} {\bibinfo  {journal} {Journal of Micromechanics and Microengineering}\ }\textbf {\bibinfo {volume} {22}},\ \bibinfo {pages} {094009} (\bibinfo {year} {2012})}\BibitemShut {NoStop}%
\bibitem [{\citenamefont {Gauster}\ and\ \citenamefont {Habing}(1967)}]{gauster1967electronic}%
  \BibitemOpen
  \bibfield  {author} {\bibinfo {author} {\bibfnamefont {W.}~\bibnamefont {Gauster}}\ and\ \bibinfo {author} {\bibfnamefont {D.}~\bibnamefont {Habing}},\ }\bibfield  {title} {\bibinfo {title} {Electronic volume effect in silicon},\ }\href@noop {} {\bibfield  {journal} {\bibinfo  {journal} {Physical Review Letters}\ }\textbf {\bibinfo {volume} {18}},\ \bibinfo {pages} {1058} (\bibinfo {year} {1967})}\BibitemShut {NoStop}%
\bibitem [{\citenamefont {Figielski}(1961)}]{figielski1961photostriction}%
  \BibitemOpen
  \bibfield  {author} {\bibinfo {author} {\bibfnamefont {T.}~\bibnamefont {Figielski}},\ }\bibfield  {title} {\bibinfo {title} {Photostriction effect in germanium},\ }\href@noop {} {\bibfield  {journal} {\bibinfo  {journal} {Physica Status Solidi (b)}\ }\textbf {\bibinfo {volume} {1}},\ \bibinfo {pages} {306} (\bibinfo {year} {1961})}\BibitemShut {NoStop}%
\bibitem [{\citenamefont {North}\ and\ \citenamefont {Buschert}(1966)}]{north1966length}%
  \BibitemOpen
  \bibfield  {author} {\bibinfo {author} {\bibfnamefont {J.}~\bibnamefont {North}}\ and\ \bibinfo {author} {\bibfnamefont {R.}~\bibnamefont {Buschert}},\ }\bibfield  {title} {\bibinfo {title} {Length changes in electron-irradiated n-and p-type germanium},\ }\href@noop {} {\bibfield  {journal} {\bibinfo  {journal} {Physical Review}\ }\textbf {\bibinfo {volume} {143}},\ \bibinfo {pages} {609} (\bibinfo {year} {1966})}\BibitemShut {NoStop}%
\bibitem [{\citenamefont {{\L}agowski}\ and\ \citenamefont {Gatos}(1974)}]{lagowski1974photomechanical}%
  \BibitemOpen
  \bibfield  {author} {\bibinfo {author} {\bibfnamefont {J.}~\bibnamefont {{\L}agowski}}\ and\ \bibinfo {author} {\bibfnamefont {H.~C.}\ \bibnamefont {Gatos}},\ }\bibfield  {title} {\bibinfo {title} {Photomechanical vibration of thin crystals of polar semiconductors},\ }\href@noop {} {\bibfield  {journal} {\bibinfo  {journal} {Surface Science}\ }\textbf {\bibinfo {volume} {45}},\ \bibinfo {pages} {353} (\bibinfo {year} {1974})}\BibitemShut {NoStop}%
\bibitem [{\citenamefont {Lagowski}\ and\ \citenamefont {Gatos}(1972)}]{lagowski1972photomechanical}%
  \BibitemOpen
  \bibfield  {author} {\bibinfo {author} {\bibfnamefont {J.}~\bibnamefont {Lagowski}}\ and\ \bibinfo {author} {\bibfnamefont {H.}~\bibnamefont {Gatos}},\ }\bibfield  {title} {\bibinfo {title} {Photomechanical effect in noncentrosymmetric semiconductors-{CdS}},\ }\href@noop {} {\bibfield  {journal} {\bibinfo  {journal} {Applied Physics Letters}\ }\textbf {\bibinfo {volume} {20}},\ \bibinfo {pages} {14} (\bibinfo {year} {1972})}\BibitemShut {NoStop}%
\bibitem [{\citenamefont {Schick}\ \emph {et~al.}(2014)\citenamefont {Schick}, \citenamefont {Herzog}, \citenamefont {Wen}, \citenamefont {Chen}, \citenamefont {Adamo}, \citenamefont {Gaal}, \citenamefont {Schlom}, \citenamefont {Evans}, \citenamefont {Li},\ and\ \citenamefont {Bargheer}}]{schick2014localized}%
  \BibitemOpen
  \bibfield  {author} {\bibinfo {author} {\bibfnamefont {D.}~\bibnamefont {Schick}}, \bibinfo {author} {\bibfnamefont {M.}~\bibnamefont {Herzog}}, \bibinfo {author} {\bibfnamefont {H.}~\bibnamefont {Wen}}, \bibinfo {author} {\bibfnamefont {P.}~\bibnamefont {Chen}}, \bibinfo {author} {\bibfnamefont {C.}~\bibnamefont {Adamo}}, \bibinfo {author} {\bibfnamefont {P.}~\bibnamefont {Gaal}}, \bibinfo {author} {\bibfnamefont {D.~G.}\ \bibnamefont {Schlom}}, \bibinfo {author} {\bibfnamefont {P.~G.}\ \bibnamefont {Evans}}, \bibinfo {author} {\bibfnamefont {Y.}~\bibnamefont {Li}},\ and\ \bibinfo {author} {\bibfnamefont {M.}~\bibnamefont {Bargheer}},\ }\bibfield  {title} {\bibinfo {title} {Localized excited charge carriers generate ultrafast inhomogeneous strain in the multiferroic {BiFeO}$_3$},\ }\href@noop {} {\bibfield  {journal} {\bibinfo  {journal} {Physical Review Letters}\ }\textbf {\bibinfo {volume} {112}},\ \bibinfo {pages} {097602} (\bibinfo {year} {2014})}\BibitemShut {NoStop}%
\bibitem [{\citenamefont {Paillard}\ \emph {et~al.}(2016)\citenamefont {Paillard}, \citenamefont {Xu}, \citenamefont {Dkhil}, \citenamefont {Geneste},\ and\ \citenamefont {Bellaiche}}]{paillard2016photostriction}%
  \BibitemOpen
  \bibfield  {author} {\bibinfo {author} {\bibfnamefont {C.}~\bibnamefont {Paillard}}, \bibinfo {author} {\bibfnamefont {B.}~\bibnamefont {Xu}}, \bibinfo {author} {\bibfnamefont {B.}~\bibnamefont {Dkhil}}, \bibinfo {author} {\bibfnamefont {G.}~\bibnamefont {Geneste}},\ and\ \bibinfo {author} {\bibfnamefont {L.}~\bibnamefont {Bellaiche}},\ }\bibfield  {title} {\bibinfo {title} {Photostriction in ferroelectrics from density functional theory},\ }\href@noop {} {\bibfield  {journal} {\bibinfo  {journal} {Physical Review Letters}\ }\textbf {\bibinfo {volume} {116}},\ \bibinfo {pages} {247401} (\bibinfo {year} {2016})}\BibitemShut {NoStop}%
\bibitem [{\citenamefont {Dogan}\ \emph {et~al.}(2001)\citenamefont {Dogan}, \citenamefont {Poosanaas}, \citenamefont {Abothu}, \citenamefont {Komarneni},\ and\ \citenamefont {Uchino}}]{dogan2001nanocomposite}%
  \BibitemOpen
  \bibfield  {author} {\bibinfo {author} {\bibfnamefont {A.}~\bibnamefont {Dogan}}, \bibinfo {author} {\bibfnamefont {P.}~\bibnamefont {Poosanaas}}, \bibinfo {author} {\bibfnamefont {I.~R.}\ \bibnamefont {Abothu}}, \bibinfo {author} {\bibfnamefont {S.}~\bibnamefont {Komarneni}},\ and\ \bibinfo {author} {\bibfnamefont {K.}~\bibnamefont {Uchino}},\ }\bibfield  {title} {\bibinfo {title} {Nanocomposite {PLZT} ceramic materials in comparison with other processing technique for photostrictive application},\ }\href@noop {} {\bibfield  {journal} {\bibinfo  {journal} {Journal of the Ceramic Society of Japan}\ }\textbf {\bibinfo {volume} {109}},\ \bibinfo {pages} {493} (\bibinfo {year} {2001})}\BibitemShut {NoStop}%
\bibitem [{\citenamefont {Kundys}(2015)}]{kundys2015photostrictive}%
  \BibitemOpen
  \bibfield  {author} {\bibinfo {author} {\bibfnamefont {B.}~\bibnamefont {Kundys}},\ }\bibfield  {title} {\bibinfo {title} {Photostrictive materials},\ }\href@noop {} {\bibfield  {journal} {\bibinfo  {journal} {Applied Physics Reviews}\ }\textbf {\bibinfo {volume} {2}} (\bibinfo {year} {2015})}\BibitemShut {NoStop}%
\bibitem [{\citenamefont {Kuzukawa}\ \emph {et~al.}(1998)\citenamefont {Kuzukawa}, \citenamefont {Ganjoo},\ and\ \citenamefont {Shimakawa}}]{kuzukawa1998photoinduced}%
  \BibitemOpen
  \bibfield  {author} {\bibinfo {author} {\bibfnamefont {Y.}~\bibnamefont {Kuzukawa}}, \bibinfo {author} {\bibfnamefont {A.}~\bibnamefont {Ganjoo}},\ and\ \bibinfo {author} {\bibfnamefont {K.}~\bibnamefont {Shimakawa}},\ }\bibfield  {title} {\bibinfo {title} {Photoinduced structural changes in obliquely deposited {As}-and {Ge}-based amorphous chalcogenides: correlation between changes in thickness and band gap},\ }\href@noop {} {\bibfield  {journal} {\bibinfo  {journal} {Journal of Non-crystalline Solids}\ }\textbf {\bibinfo {volume} {227}},\ \bibinfo {pages} {715} (\bibinfo {year} {1998})}\BibitemShut {NoStop}%
\bibitem [{\citenamefont {Finkelmann}\ \emph {et~al.}(2001)\citenamefont {Finkelmann}, \citenamefont {Nishikawa}, \citenamefont {Pereira},\ and\ \citenamefont {Warner}}]{finkelmann2001new}%
  \BibitemOpen
  \bibfield  {author} {\bibinfo {author} {\bibfnamefont {H.}~\bibnamefont {Finkelmann}}, \bibinfo {author} {\bibfnamefont {E.}~\bibnamefont {Nishikawa}}, \bibinfo {author} {\bibfnamefont {G.}~\bibnamefont {Pereira}},\ and\ \bibinfo {author} {\bibfnamefont {M.}~\bibnamefont {Warner}},\ }\bibfield  {title} {\bibinfo {title} {A new opto-mechanical effect in solids},\ }\href@noop {} {\bibfield  {journal} {\bibinfo  {journal} {Physical Review Letters}\ }\textbf {\bibinfo {volume} {87}},\ \bibinfo {pages} {015501} (\bibinfo {year} {2001})}\BibitemShut {NoStop}%
\bibitem [{\citenamefont {Zhou}\ \emph {et~al.}(2016)\citenamefont {Zhou}, \citenamefont {You}, \citenamefont {Wang}, \citenamefont {Ku}, \citenamefont {Fan}, \citenamefont {Schmidt}, \citenamefont {Rusydi}, \citenamefont {Chang}, \citenamefont {Wang}, \citenamefont {Ren} \emph {et~al.}}]{zhou2016giant}%
  \BibitemOpen
  \bibfield  {author} {\bibinfo {author} {\bibfnamefont {Y.}~\bibnamefont {Zhou}}, \bibinfo {author} {\bibfnamefont {L.}~\bibnamefont {You}}, \bibinfo {author} {\bibfnamefont {S.}~\bibnamefont {Wang}}, \bibinfo {author} {\bibfnamefont {Z.}~\bibnamefont {Ku}}, \bibinfo {author} {\bibfnamefont {H.}~\bibnamefont {Fan}}, \bibinfo {author} {\bibfnamefont {D.}~\bibnamefont {Schmidt}}, \bibinfo {author} {\bibfnamefont {A.}~\bibnamefont {Rusydi}}, \bibinfo {author} {\bibfnamefont {L.}~\bibnamefont {Chang}}, \bibinfo {author} {\bibfnamefont {L.}~\bibnamefont {Wang}}, \bibinfo {author} {\bibfnamefont {P.}~\bibnamefont {Ren}}, \emph {et~al.},\ }\bibfield  {title} {\bibinfo {title} {Giant photostriction in organic--inorganic lead halide perovskites},\ }\href@noop {} {\bibfield  {journal} {\bibinfo  {journal} {Nature Communications}\ }\textbf {\bibinfo {volume} {7}},\ \bibinfo {pages} {11193} (\bibinfo {year} {2016})}\BibitemShut {NoStop}%
\bibitem [{\citenamefont {Lv}\ \emph {et~al.}(2021)\citenamefont {Lv}, \citenamefont {Dong}, \citenamefont {Huang}, \citenamefont {Cao}, \citenamefont {Zeng}, \citenamefont {Wang}, \citenamefont {Wu}, \citenamefont {Chen}, \citenamefont {Wang}, \citenamefont {Yuan} \emph {et~al.}}]{lv2021giant}%
  \BibitemOpen
  \bibfield  {author} {\bibinfo {author} {\bibfnamefont {X.}~\bibnamefont {Lv}}, \bibinfo {author} {\bibfnamefont {S.}~\bibnamefont {Dong}}, \bibinfo {author} {\bibfnamefont {X.}~\bibnamefont {Huang}}, \bibinfo {author} {\bibfnamefont {B.}~\bibnamefont {Cao}}, \bibinfo {author} {\bibfnamefont {S.}~\bibnamefont {Zeng}}, \bibinfo {author} {\bibfnamefont {Y.}~\bibnamefont {Wang}}, \bibinfo {author} {\bibfnamefont {T.}~\bibnamefont {Wu}}, \bibinfo {author} {\bibfnamefont {L.}~\bibnamefont {Chen}}, \bibinfo {author} {\bibfnamefont {J.}~\bibnamefont {Wang}}, \bibinfo {author} {\bibfnamefont {G.}~\bibnamefont {Yuan}}, \emph {et~al.},\ }\bibfield  {title} {\bibinfo {title} {Giant bulk photostriction and accurate photomechanical actuation in hybrid perovskites},\ }\href@noop {} {\bibfield  {journal} {\bibinfo  {journal} {Advanced Optical Materials}\ }\textbf {\bibinfo {volume} {9}},\ \bibinfo {pages} {2100837} (\bibinfo {year} {2021})}\BibitemShut {NoStop}%
\bibitem [{\citenamefont {Wei}\ \emph {et~al.}(2017)\citenamefont {Wei}, \citenamefont {Wang}, \citenamefont {Liu}, \citenamefont {Tsai}, \citenamefont {Ke}, \citenamefont {Wu}, \citenamefont {Yin}, \citenamefont {Zhan}, \citenamefont {Lin}, \citenamefont {Chu} \emph {et~al.}}]{wei2017photostriction}%
  \BibitemOpen
  \bibfield  {author} {\bibinfo {author} {\bibfnamefont {T.-C.}\ \bibnamefont {Wei}}, \bibinfo {author} {\bibfnamefont {H.-P.}\ \bibnamefont {Wang}}, \bibinfo {author} {\bibfnamefont {H.-J.}\ \bibnamefont {Liu}}, \bibinfo {author} {\bibfnamefont {D.-S.}\ \bibnamefont {Tsai}}, \bibinfo {author} {\bibfnamefont {J.-J.}\ \bibnamefont {Ke}}, \bibinfo {author} {\bibfnamefont {C.-L.}\ \bibnamefont {Wu}}, \bibinfo {author} {\bibfnamefont {Y.-P.}\ \bibnamefont {Yin}}, \bibinfo {author} {\bibfnamefont {Q.}~\bibnamefont {Zhan}}, \bibinfo {author} {\bibfnamefont {G.-R.}\ \bibnamefont {Lin}}, \bibinfo {author} {\bibfnamefont {Y.-H.}\ \bibnamefont {Chu}}, \emph {et~al.},\ }\bibfield  {title} {\bibinfo {title} {Photostriction of strontium ruthenate},\ }\href@noop {} {\bibfield  {journal} {\bibinfo  {journal} {Nature Communications}\ }\textbf {\bibinfo {volume} {8}},\ \bibinfo {pages} {15018} (\bibinfo {year} {2017})}\BibitemShut {NoStop}%
\bibitem [{\citenamefont {Kaduk}\ \emph {et~al.}(2012)\citenamefont {Kaduk}, \citenamefont {Kowalczyk},\ and\ \citenamefont {Van~Voorhis}}]{kaduk2012constrained}%
  \BibitemOpen
  \bibfield  {author} {\bibinfo {author} {\bibfnamefont {B.}~\bibnamefont {Kaduk}}, \bibinfo {author} {\bibfnamefont {T.}~\bibnamefont {Kowalczyk}},\ and\ \bibinfo {author} {\bibfnamefont {T.}~\bibnamefont {Van~Voorhis}},\ }\bibfield  {title} {\bibinfo {title} {Constrained density functional theory},\ }\href@noop {} {\bibfield  {journal} {\bibinfo  {journal} {Chemical Reviews}\ }\textbf {\bibinfo {volume} {112}},\ \bibinfo {pages} {321} (\bibinfo {year} {2012})}\BibitemShut {NoStop}%
\bibitem [{\citenamefont {Paillard}\ \emph {et~al.}(2017)\citenamefont {Paillard}, \citenamefont {Prosandeev},\ and\ \citenamefont {Bellaiche}}]{paillard2017ab}%
  \BibitemOpen
  \bibfield  {author} {\bibinfo {author} {\bibfnamefont {C.}~\bibnamefont {Paillard}}, \bibinfo {author} {\bibfnamefont {S.}~\bibnamefont {Prosandeev}},\ and\ \bibinfo {author} {\bibfnamefont {L.}~\bibnamefont {Bellaiche}},\ }\bibfield  {title} {\bibinfo {title} {Ab initio approach to photostriction in classical ferroelectric materials},\ }\href@noop {} {\bibfield  {journal} {\bibinfo  {journal} {Physical Review B}\ }\textbf {\bibinfo {volume} {96}},\ \bibinfo {pages} {045205} (\bibinfo {year} {2017})}\BibitemShut {NoStop}%
\bibitem [{\citenamefont {Peng}\ \emph {et~al.}(2022)\citenamefont {Peng}, \citenamefont {Bennett}, \citenamefont {Bravi{\'c}},\ and\ \citenamefont {Monserrat}}]{peng2022tunable}%
  \BibitemOpen
  \bibfield  {author} {\bibinfo {author} {\bibfnamefont {B.}~\bibnamefont {Peng}}, \bibinfo {author} {\bibfnamefont {D.}~\bibnamefont {Bennett}}, \bibinfo {author} {\bibfnamefont {I.}~\bibnamefont {Bravi{\'c}}},\ and\ \bibinfo {author} {\bibfnamefont {B.}~\bibnamefont {Monserrat}},\ }\bibfield  {title} {\bibinfo {title} {Tunable photostriction of halide perovskites through energy dependent photoexcitation},\ }\href@noop {} {\bibfield  {journal} {\bibinfo  {journal} {Physical Review Materials}\ }\textbf {\bibinfo {volume} {6}},\ \bibinfo {pages} {L082401} (\bibinfo {year} {2022})}\BibitemShut {NoStop}%
\bibitem [{\citenamefont {Dronskowski}\ and\ \citenamefont {Bloechl}(1993)}]{dronskowski1993crystal}%
  \BibitemOpen
  \bibfield  {author} {\bibinfo {author} {\bibfnamefont {R.}~\bibnamefont {Dronskowski}}\ and\ \bibinfo {author} {\bibfnamefont {P.~E.}\ \bibnamefont {Bloechl}},\ }\bibfield  {title} {\bibinfo {title} {Crystal orbital {Hamilton} populations {(COHP):} {Energy-resolved} visualization of chemical bonding in solids based on density-functional calculations},\ }\href {https://doi.org/10.1021/j100135a014} {\bibfield  {journal} {\bibinfo  {journal} {The Journal of Physical Chemistry}\ }\textbf {\bibinfo {volume} {97}},\ \bibinfo {pages} {8617} (\bibinfo {year} {1993})}\BibitemShut {NoStop}%
\bibitem [{\citenamefont {Deringer}\ \emph {et~al.}(2011)\citenamefont {Deringer}, \citenamefont {Tchougr\'eeff},\ and\ \citenamefont {Dronskowski}}]{deringer2011crystal}%
  \BibitemOpen
  \bibfield  {author} {\bibinfo {author} {\bibfnamefont {V.~L.}\ \bibnamefont {Deringer}}, \bibinfo {author} {\bibfnamefont {A.~L.}\ \bibnamefont {Tchougr\'eeff}},\ and\ \bibinfo {author} {\bibfnamefont {R.}~\bibnamefont {Dronskowski}},\ }\bibfield  {title} {\bibinfo {title} {Crystal orbital {Hamilton} population {(COHP)} analysis as projected from plane-wave basis sets},\ }\href {https://doi.org/10.1021/jp202489s} {\bibfield  {journal} {\bibinfo  {journal} {J. Phys. Chem. A}\ }\textbf {\bibinfo {volume} {115}},\ \bibinfo {pages} {5461} (\bibinfo {year} {2011})}\BibitemShut {NoStop}%
\bibitem [{\citenamefont {W{\"u}rfel}\ and\ \citenamefont {W{\"u}rfel}(2016)}]{wurfel2016physics}%
  \BibitemOpen
  \bibfield  {author} {\bibinfo {author} {\bibfnamefont {P.}~\bibnamefont {W{\"u}rfel}}\ and\ \bibinfo {author} {\bibfnamefont {U.}~\bibnamefont {W{\"u}rfel}},\ }\href@noop {} {\emph {\bibinfo {title} {Physics of solar cells: from basic principles to advanced concepts}}}\ (\bibinfo  {publisher} {John Wiley \& Sons},\ \bibinfo {year} {2016})\BibitemShut {NoStop}%
\bibitem [{\citenamefont {Kresse}\ and\ \citenamefont {Furthm{\"u}ller}(1996{\natexlab{a}})}]{kresse1996efficiency}%
  \BibitemOpen
  \bibfield  {author} {\bibinfo {author} {\bibfnamefont {G.}~\bibnamefont {Kresse}}\ and\ \bibinfo {author} {\bibfnamefont {J.}~\bibnamefont {Furthm{\"u}ller}},\ }\bibfield  {title} {\bibinfo {title} {Efficiency of ab-initio total energy calculations for metals and semiconductors using a plane-wave basis set},\ }\href@noop {} {\bibfield  {journal} {\bibinfo  {journal} {Computational Materials Science}\ }\textbf {\bibinfo {volume} {6}},\ \bibinfo {pages} {15} (\bibinfo {year} {1996}{\natexlab{a}})}\BibitemShut {NoStop}%
\bibitem [{\citenamefont {Kresse}\ and\ \citenamefont {Furthm{\"u}ller}(1996{\natexlab{b}})}]{kresse1996efficient}%
  \BibitemOpen
  \bibfield  {author} {\bibinfo {author} {\bibfnamefont {G.}~\bibnamefont {Kresse}}\ and\ \bibinfo {author} {\bibfnamefont {J.}~\bibnamefont {Furthm{\"u}ller}},\ }\bibfield  {title} {\bibinfo {title} {Efficient iterative schemes for ab initio total-energy calculations using a plane-wave basis set},\ }\href@noop {} {\bibfield  {journal} {\bibinfo  {journal} {Physical Review B}\ }\textbf {\bibinfo {volume} {54}},\ \bibinfo {pages} {11169} (\bibinfo {year} {1996}{\natexlab{b}})}\BibitemShut {NoStop}%
\bibitem [{\citenamefont {Bl{\"o}chl}(1994)}]{blochl1994projector}%
  \BibitemOpen
  \bibfield  {author} {\bibinfo {author} {\bibfnamefont {P.~E.}\ \bibnamefont {Bl{\"o}chl}},\ }\bibfield  {title} {\bibinfo {title} {Projector augmented-wave method},\ }\href@noop {} {\bibfield  {journal} {\bibinfo  {journal} {Physical Review B}\ }\textbf {\bibinfo {volume} {50}},\ \bibinfo {pages} {17953} (\bibinfo {year} {1994})}\BibitemShut {NoStop}%
\bibitem [{\citenamefont {Perdew}\ \emph {et~al.}(1996)\citenamefont {Perdew}, \citenamefont {Burke},\ and\ \citenamefont {Ernzerhof}}]{perdew1996generalized}%
  \BibitemOpen
  \bibfield  {author} {\bibinfo {author} {\bibfnamefont {J.~P.}\ \bibnamefont {Perdew}}, \bibinfo {author} {\bibfnamefont {K.}~\bibnamefont {Burke}},\ and\ \bibinfo {author} {\bibfnamefont {M.}~\bibnamefont {Ernzerhof}},\ }\bibfield  {title} {\bibinfo {title} {Generalized gradient approximation made simple},\ }\href@noop {} {\bibfield  {journal} {\bibinfo  {journal} {Physical Review Letters}\ }\textbf {\bibinfo {volume} {77}},\ \bibinfo {pages} {3865} (\bibinfo {year} {1996})}\BibitemShut {NoStop}%
\bibitem [{\citenamefont {Jain}\ \emph {et~al.}(2013)\citenamefont {Jain}, \citenamefont {Ong}, \citenamefont {Hautier}, \citenamefont {Chen}, \citenamefont {Richards}, \citenamefont {Dacek}, \citenamefont {Cholia}, \citenamefont {Gunter}, \citenamefont {Skinner}, \citenamefont {Ceder} \emph {et~al.}}]{jain2013commentary}%
  \BibitemOpen
  \bibfield  {author} {\bibinfo {author} {\bibfnamefont {A.}~\bibnamefont {Jain}}, \bibinfo {author} {\bibfnamefont {S.~P.}\ \bibnamefont {Ong}}, \bibinfo {author} {\bibfnamefont {G.}~\bibnamefont {Hautier}}, \bibinfo {author} {\bibfnamefont {W.}~\bibnamefont {Chen}}, \bibinfo {author} {\bibfnamefont {W.~D.}\ \bibnamefont {Richards}}, \bibinfo {author} {\bibfnamefont {S.}~\bibnamefont {Dacek}}, \bibinfo {author} {\bibfnamefont {S.}~\bibnamefont {Cholia}}, \bibinfo {author} {\bibfnamefont {D.}~\bibnamefont {Gunter}}, \bibinfo {author} {\bibfnamefont {D.}~\bibnamefont {Skinner}}, \bibinfo {author} {\bibfnamefont {G.}~\bibnamefont {Ceder}}, \emph {et~al.},\ }\bibfield  {title} {\bibinfo {title} {Commentary: The materials project: A materials genome approach to accelerating materials innovation},\ }\href@noop {} {\bibfield  {journal} {\bibinfo  {journal} {APL Materials}\ }\textbf {\bibinfo {volume} {1}} (\bibinfo {year} {2013})}\BibitemShut {NoStop}%
\bibitem [{\citenamefont {Welber}\ \emph {et~al.}(1975)\citenamefont {Welber}, \citenamefont {Cardona}, \citenamefont {Kim},\ and\ \citenamefont {Rodriguez}}]{welber1975dependence}%
  \BibitemOpen
  \bibfield  {author} {\bibinfo {author} {\bibfnamefont {B.}~\bibnamefont {Welber}}, \bibinfo {author} {\bibfnamefont {M.}~\bibnamefont {Cardona}}, \bibinfo {author} {\bibfnamefont {C.}~\bibnamefont {Kim}},\ and\ \bibinfo {author} {\bibfnamefont {S.}~\bibnamefont {Rodriguez}},\ }\bibfield  {title} {\bibinfo {title} {Dependence of the direct energy gap of gaas on hydrostatic pressure},\ }\href@noop {} {\bibfield  {journal} {\bibinfo  {journal} {Physical Review B}\ }\textbf {\bibinfo {volume} {12}},\ \bibinfo {pages} {5729} (\bibinfo {year} {1975})}\BibitemShut {NoStop}%
\bibitem [{\citenamefont {Wei}\ and\ \citenamefont {Zunger}(1999)}]{wei1999predicted}%
  \BibitemOpen
  \bibfield  {author} {\bibinfo {author} {\bibfnamefont {S.-H.}\ \bibnamefont {Wei}}\ and\ \bibinfo {author} {\bibfnamefont {A.}~\bibnamefont {Zunger}},\ }\bibfield  {title} {\bibinfo {title} {Predicted band-gap pressure coefficients of all diamond and zinc-blende semiconductors: Chemical trends},\ }\href@noop {} {\bibfield  {journal} {\bibinfo  {journal} {Physical Review B}\ }\textbf {\bibinfo {volume} {60}},\ \bibinfo {pages} {5404} (\bibinfo {year} {1999})}\BibitemShut {NoStop}%
\bibitem [{\citenamefont {Dornhaus}\ \emph {et~al.}(2006)\citenamefont {Dornhaus}, \citenamefont {Nimtz},\ and\ \citenamefont {Schlicht}}]{dornhaus2006narrow}%
  \BibitemOpen
  \bibfield  {author} {\bibinfo {author} {\bibfnamefont {R.}~\bibnamefont {Dornhaus}}, \bibinfo {author} {\bibfnamefont {G.}~\bibnamefont {Nimtz}},\ and\ \bibinfo {author} {\bibfnamefont {B.}~\bibnamefont {Schlicht}},\ }\href@noop {} {\emph {\bibinfo {title} {Narrow-gap semiconductors}}},\ Vol.~\bibinfo {volume} {98}\ (\bibinfo  {publisher} {Springer},\ \bibinfo {year} {2006})\BibitemShut {NoStop}%
\bibitem [{\citenamefont {Wei}\ and\ \citenamefont {Zunger}(1997)}]{wei1997electronic}%
  \BibitemOpen
  \bibfield  {author} {\bibinfo {author} {\bibfnamefont {S.-H.}\ \bibnamefont {Wei}}\ and\ \bibinfo {author} {\bibfnamefont {A.}~\bibnamefont {Zunger}},\ }\bibfield  {title} {\bibinfo {title} {Electronic and structural anomalies in lead chalcogenides},\ }\href@noop {} {\bibfield  {journal} {\bibinfo  {journal} {Physical Review B}\ }\textbf {\bibinfo {volume} {55}},\ \bibinfo {pages} {13605} (\bibinfo {year} {1997})}\BibitemShut {NoStop}%
\bibitem [{\citenamefont {Yuan}\ \emph {et~al.}(2023)\citenamefont {Yuan}, \citenamefont {Chen},\ and\ \citenamefont {Liao}}]{yuan2023lattice}%
  \BibitemOpen
  \bibfield  {author} {\bibinfo {author} {\bibfnamefont {J.}~\bibnamefont {Yuan}}, \bibinfo {author} {\bibfnamefont {Y.}~\bibnamefont {Chen}},\ and\ \bibinfo {author} {\bibfnamefont {B.}~\bibnamefont {Liao}},\ }\bibfield  {title} {\bibinfo {title} {Lattice dynamics and thermal transport in semiconductors with anti-bonding valence bands},\ }\href@noop {} {\bibfield  {journal} {\bibinfo  {journal} {Journal of the American Chemical Society}\ }\textbf {\bibinfo {volume} {145}},\ \bibinfo {pages} {18506} (\bibinfo {year} {2023})}\BibitemShut {NoStop}%
\bibitem [{\citenamefont {Cheng}\ \emph {et~al.}(2019)\citenamefont {Cheng}, \citenamefont {Shulumba},\ and\ \citenamefont {Minnich}}]{cheng2019thermal}%
  \BibitemOpen
  \bibfield  {author} {\bibinfo {author} {\bibfnamefont {P.}~\bibnamefont {Cheng}}, \bibinfo {author} {\bibfnamefont {N.}~\bibnamefont {Shulumba}},\ and\ \bibinfo {author} {\bibfnamefont {A.~J.}\ \bibnamefont {Minnich}},\ }\bibfield  {title} {\bibinfo {title} {Thermal transport and phonon focusing in complex molecular crystals: {Ab} initio study of polythiophene},\ }\href@noop {} {\bibfield  {journal} {\bibinfo  {journal} {Physical Review B}\ }\textbf {\bibinfo {volume} {100}},\ \bibinfo {pages} {094306} (\bibinfo {year} {2019})}\BibitemShut {NoStop}%
\bibitem [{\citenamefont {Wang}\ \emph {et~al.}(2015)\citenamefont {Wang}, \citenamefont {Lü}, \citenamefont {Yang}, \citenamefont {Wen}, \citenamefont {Yang}, \citenamefont {Ren}, \citenamefont {Wang}, \citenamefont {Lin},\ and\ \citenamefont {Zhao}}]{wang2015pressure}%
  \BibitemOpen
  \bibfield  {author} {\bibinfo {author} {\bibfnamefont {Y.}~\bibnamefont {Wang}}, \bibinfo {author} {\bibfnamefont {X.}~\bibnamefont {Lü}}, \bibinfo {author} {\bibfnamefont {W.}~\bibnamefont {Yang}}, \bibinfo {author} {\bibfnamefont {T.}~\bibnamefont {Wen}}, \bibinfo {author} {\bibfnamefont {L.}~\bibnamefont {Yang}}, \bibinfo {author} {\bibfnamefont {X.}~\bibnamefont {Ren}}, \bibinfo {author} {\bibfnamefont {L.}~\bibnamefont {Wang}}, \bibinfo {author} {\bibfnamefont {Z.}~\bibnamefont {Lin}},\ and\ \bibinfo {author} {\bibfnamefont {Y.}~\bibnamefont {Zhao}},\ }\bibfield  {title} {\bibinfo {title} {Pressure-induced phase transformation, reversible amorphization, and anomalous visible light response in organolead bromide perovskite},\ }\href@noop {} {\bibfield  {journal} {\bibinfo  {journal} {Journal of the American Chemical Society}\ }\textbf {\bibinfo {volume} {137}},\ \bibinfo {pages} {11144} (\bibinfo {year} {2015})}\BibitemShut {NoStop}%
\bibitem [{\citenamefont {Bagri}\ \emph {et~al.}(2022)\citenamefont {Bagri}, \citenamefont {Jana}, \citenamefont {Panchal}, \citenamefont {Phase},\ and\ \citenamefont {Choudhary}}]{bagri2022amalgamation}%
  \BibitemOpen
  \bibfield  {author} {\bibinfo {author} {\bibfnamefont {A.}~\bibnamefont {Bagri}}, \bibinfo {author} {\bibfnamefont {A.}~\bibnamefont {Jana}}, \bibinfo {author} {\bibfnamefont {G.}~\bibnamefont {Panchal}}, \bibinfo {author} {\bibfnamefont {D.~M.}\ \bibnamefont {Phase}},\ and\ \bibinfo {author} {\bibfnamefont {R.~J.}\ \bibnamefont {Choudhary}},\ }\bibfield  {title} {\bibinfo {title} {Amalgamation of photostriction, photodomain, and photopolarization effects in {BaTiO$_3$} and its electronic origin},\ }\href@noop {} {\bibfield  {journal} {\bibinfo  {journal} {ACS Applied Electronic Materials}\ }\textbf {\bibinfo {volume} {4}},\ \bibinfo {pages} {4438} (\bibinfo {year} {2022})}\BibitemShut {NoStop}%
\bibitem [{\citenamefont {Soled}\ \emph {et~al.}(1976)\citenamefont {Soled}, \citenamefont {Wold},\ and\ \citenamefont {Gorochov}}]{soled1976crystal}%
  \BibitemOpen
  \bibfield  {author} {\bibinfo {author} {\bibfnamefont {S.}~\bibnamefont {Soled}}, \bibinfo {author} {\bibfnamefont {A.}~\bibnamefont {Wold}},\ and\ \bibinfo {author} {\bibfnamefont {O.}~\bibnamefont {Gorochov}},\ }\bibfield  {title} {\bibinfo {title} {Crystal growth and characterization of several platinum sulfoselenides},\ }\href@noop {} {\bibfield  {journal} {\bibinfo  {journal} {Materials Research Bulletin}\ }\textbf {\bibinfo {volume} {11}},\ \bibinfo {pages} {927} (\bibinfo {year} {1976})}\BibitemShut {NoStop}%
\bibitem [{\citenamefont {Guo}\ and\ \citenamefont {Liang}(1986)}]{guo1986electronic}%
  \BibitemOpen
  \bibfield  {author} {\bibinfo {author} {\bibfnamefont {G.}~\bibnamefont {Guo}}\ and\ \bibinfo {author} {\bibfnamefont {W.}~\bibnamefont {Liang}},\ }\bibfield  {title} {\bibinfo {title} {The electronic structures of platinum dichalcogenides: Pts2, ptse2 and ptte2},\ }\href@noop {} {\bibfield  {journal} {\bibinfo  {journal} {Journal of Physics C: Solid State Physics}\ }\textbf {\bibinfo {volume} {19}},\ \bibinfo {pages} {995} (\bibinfo {year} {1986})}\BibitemShut {NoStop}%
\bibitem [{\citenamefont {Cullen}\ \emph {et~al.}(2021)\citenamefont {Cullen}, \citenamefont {Coile{\'a}in}, \citenamefont {McManus}, \citenamefont {Hartwig}, \citenamefont {McCloskey}, \citenamefont {Duesberg},\ and\ \citenamefont {McEvoy}}]{cullen2021synthesis}%
  \BibitemOpen
  \bibfield  {author} {\bibinfo {author} {\bibfnamefont {C.~P.}\ \bibnamefont {Cullen}}, \bibinfo {author} {\bibfnamefont {C.~{\'O}.}\ \bibnamefont {Coile{\'a}in}}, \bibinfo {author} {\bibfnamefont {J.~B.}\ \bibnamefont {McManus}}, \bibinfo {author} {\bibfnamefont {O.}~\bibnamefont {Hartwig}}, \bibinfo {author} {\bibfnamefont {D.}~\bibnamefont {McCloskey}}, \bibinfo {author} {\bibfnamefont {G.~S.}\ \bibnamefont {Duesberg}},\ and\ \bibinfo {author} {\bibfnamefont {N.}~\bibnamefont {McEvoy}},\ }\bibfield  {title} {\bibinfo {title} {Synthesis and characterisation of thin-film platinum disulfide and platinum sulfide},\ }\href@noop {} {\bibfield  {journal} {\bibinfo  {journal} {Nanoscale}\ }\textbf {\bibinfo {volume} {13}},\ \bibinfo {pages} {7403} (\bibinfo {year} {2021})}\BibitemShut {NoStop}%
\bibitem [{\citenamefont {Anastassakis}\ \emph {et~al.}(1985)\citenamefont {Anastassakis}, \citenamefont {Raptis},\ and\ \citenamefont {Richter}}]{anastassakis1985raman}%
  \BibitemOpen
  \bibfield  {author} {\bibinfo {author} {\bibfnamefont {E.}~\bibnamefont {Anastassakis}}, \bibinfo {author} {\bibfnamefont {J.}~\bibnamefont {Raptis}},\ and\ \bibinfo {author} {\bibfnamefont {W.}~\bibnamefont {Richter}},\ }\bibfield  {title} {\bibinfo {title} {Raman and infrared spectra of tellurium sub-bromide (te2br)},\ }\href@noop {} {\bibfield  {journal} {\bibinfo  {journal} {Physica Status Solidi (b)}\ }\textbf {\bibinfo {volume} {130}},\ \bibinfo {pages} {161} (\bibinfo {year} {1985})}\BibitemShut {NoStop}%
\bibitem [{\citenamefont {Kniep}\ \emph {et~al.}(1976)\citenamefont {Kniep}, \citenamefont {Mootz},\ and\ \citenamefont {Rabenau}}]{kniep1976kenntnis}%
  \BibitemOpen
  \bibfield  {author} {\bibinfo {author} {\bibfnamefont {R.}~\bibnamefont {Kniep}}, \bibinfo {author} {\bibfnamefont {D.}~\bibnamefont {Mootz}},\ and\ \bibinfo {author} {\bibfnamefont {A.}~\bibnamefont {Rabenau}},\ }\bibfield  {title} {\bibinfo {title} {Zur kenntnis der subhalogenide des tellurs},\ }\href@noop {} {\bibfield  {journal} {\bibinfo  {journal} {Zeitschrift f{\"u}r anorganische und allgemeine Chemie}\ }\textbf {\bibinfo {volume} {422}},\ \bibinfo {pages} {17} (\bibinfo {year} {1976})}\BibitemShut {NoStop}%
\bibitem [{\citenamefont {Harrison}(2012)}]{harrison2012electronic}%
  \BibitemOpen
  \bibfield  {author} {\bibinfo {author} {\bibfnamefont {W.~A.}\ \bibnamefont {Harrison}},\ }\href@noop {} {\emph {\bibinfo {title} {Electronic structure and the properties of solids: the physics of the chemical bond}}}\ (\bibinfo  {publisher} {Courier Corporation},\ \bibinfo {year} {2012})\BibitemShut {NoStop}%
\end{thebibliography}%


\begin{thebibliography}{1}%
\makeatletter
\providecommand \@ifxundefined [1]{%
 \@ifx{#1\undefined}
}%
\providecommand \@ifnum [1]{%
 \ifnum #1\expandafter \@firstoftwo
 \else \expandafter \@secondoftwo
 \fi
}%
\providecommand \@ifx [1]{%
 \ifx #1\expandafter \@firstoftwo
 \else \expandafter \@secondoftwo
 \fi
}%
\providecommand \natexlab [1]{#1}%
\providecommand \enquote  [1]{``#1''}%
\providecommand \bibnamefont  [1]{#1}%
\providecommand \bibfnamefont [1]{#1}%
\providecommand \citenamefont [1]{#1}%
\providecommand \href@noop [0]{\@secondoftwo}%
\providecommand \href [0]{\begingroup \@sanitize@url \@href}%
\providecommand \@href[1]{\@@startlink{#1}\@@href}%
\providecommand \@@href[1]{\endgroup#1\@@endlink}%
\providecommand \@sanitize@url [0]{\catcode `\\12\catcode `\$12\catcode `\&12\catcode `\#12\catcode `\^12\catcode `\_12\catcode `\%12\relax}%
\providecommand \@@startlink[1]{}%
\providecommand \@@endlink[0]{}%
\providecommand \url  [0]{\begingroup\@sanitize@url \@url }%
\providecommand \@url [1]{\endgroup\@href {#1}{\urlprefix }}%
\providecommand \urlprefix  [0]{URL }%
\providecommand \Eprint [0]{\href }%
\providecommand \doibase [0]{https://doi.org/}%
\providecommand \selectlanguage [0]{\@gobble}%
\providecommand \bibinfo  [0]{\@secondoftwo}%
\providecommand \bibfield  [0]{\@secondoftwo}%
\providecommand \translation [1]{[#1]}%
\providecommand \BibitemOpen [0]{}%
\providecommand \bibitemStop [0]{}%
\providecommand \bibitemNoStop [0]{.\EOS\space}%
\providecommand \EOS [0]{\spacefactor3000\relax}%
\providecommand \BibitemShut  [1]{\csname bibitem#1\endcsname}%
\let\auto@bib@innerbib\@empty
\bibitem [{\citenamefont {Jain}\ \emph {et~al.}(2013)\citenamefont {Jain}, \citenamefont {Ong}, \citenamefont {Hautier}, \citenamefont {Chen}, \citenamefont {Richards}, \citenamefont {Dacek}, \citenamefont {Cholia}, \citenamefont {Gunter}, \citenamefont {Skinner}, \citenamefont {Ceder} \emph {et~al.}}]{jain2013commentary}%
  \BibitemOpen
  \bibfield  {author} {\bibinfo {author} {\bibfnamefont {A.}~\bibnamefont {Jain}}, \bibinfo {author} {\bibfnamefont {S.~P.}\ \bibnamefont {Ong}}, \bibinfo {author} {\bibfnamefont {G.}~\bibnamefont {Hautier}}, \bibinfo {author} {\bibfnamefont {W.}~\bibnamefont {Chen}}, \bibinfo {author} {\bibfnamefont {W.~D.}\ \bibnamefont {Richards}}, \bibinfo {author} {\bibfnamefont {S.}~\bibnamefont {Dacek}}, \bibinfo {author} {\bibfnamefont {S.}~\bibnamefont {Cholia}}, \bibinfo {author} {\bibfnamefont {D.}~\bibnamefont {Gunter}}, \bibinfo {author} {\bibfnamefont {D.}~\bibnamefont {Skinner}}, \bibinfo {author} {\bibfnamefont {G.}~\bibnamefont {Ceder}}, \emph {et~al.},\ }\bibfield  {title} {\bibinfo {title} {Commentary: The materials project: A materials genome approach to accelerating materials innovation},\ }\href@noop {} {\bibfield  {journal} {\bibinfo  {journal} {APL Materials}\ }\textbf {\bibinfo {volume} {1}} (\bibinfo {year} {2013})}\BibitemShut {NoStop}%
\end{thebibliography}%

\end{document}



\title{Supplementary Information: High-Throughput Search for Photostrictive Materials based on a Thermodynamic Descriptor} 

\author{Zeyu Xiang}
\thanks{These authors contributed equally.}
\affiliation{Department of Mechanical Engineering, University of California, Santa Barbara, CA 93106, USA}

\author{Yubi Chen}
\thanks{These authors contributed equally.}
\affiliation{Department of Mechanical Engineering, University of California, Santa Barbara, CA 93106, USA}
\affiliation{Department of Physics, University of California, Santa Barbara, CA 93106, USA}

\author{Yujie Quan}
\affiliation{Department of Mechanical Engineering, University of California, Santa Barbara, CA 93106, USA}

\author{Bolin Liao}
\email{bliao@ucsb.edu} \affiliation{Department of Mechanical Engineering, University of California, Santa Barbara, CA 93106, USA}

\date{\today}

\maketitle



\section{Lattice Structures And Crystal Orientations}

\subsection{GaN, MAPbI$_3$, PT, and BaTiO$_3$}
\begin{figure}[!htb]
\includegraphics[width=
\textwidth]{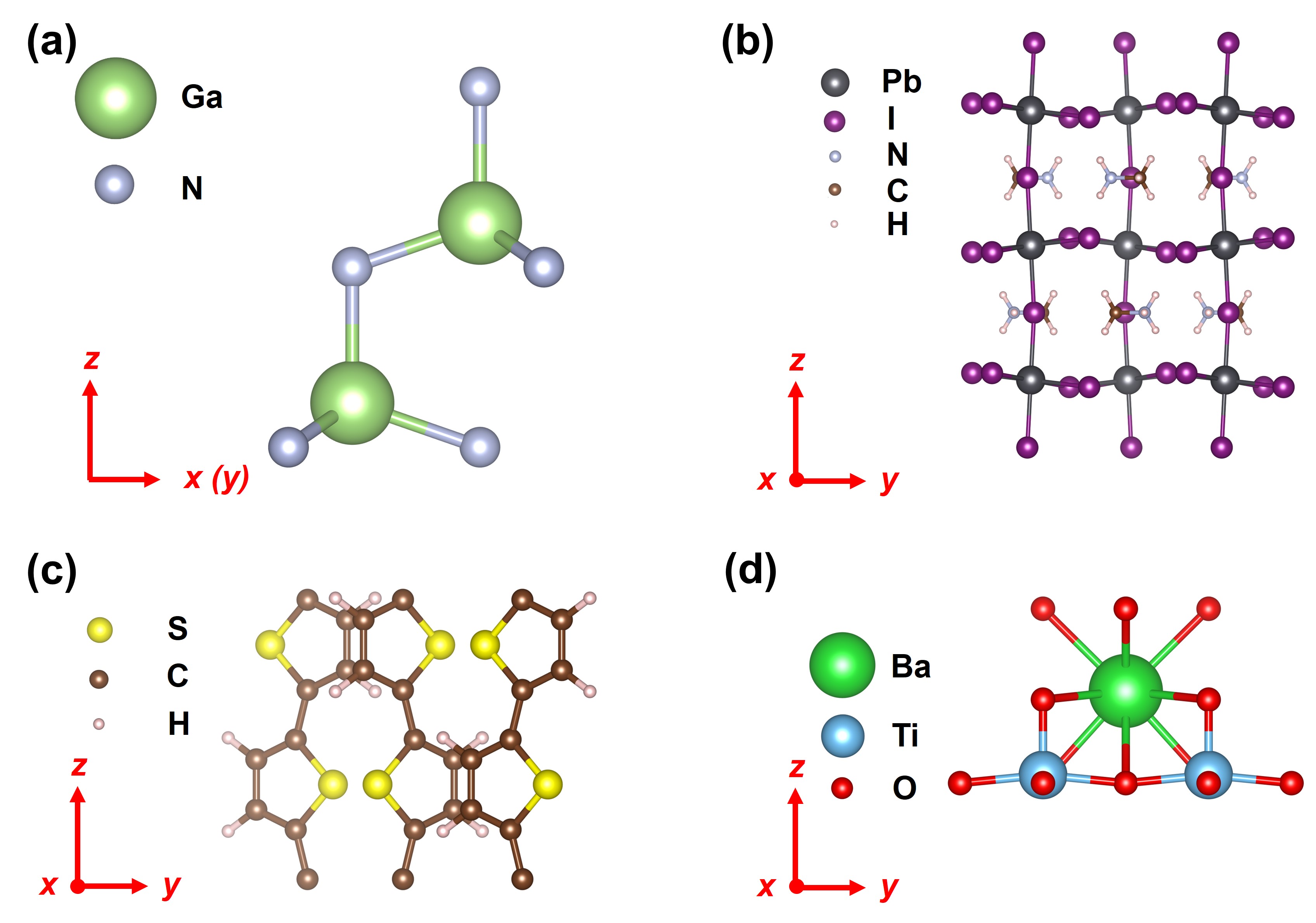}
\caption{Schematic illustrations of lattice structures and crystal orientations for the unit cell of several anisotropic materials studied in this work. 
\textbf{a}, GaN. 
\textbf{b}, MAPbI$_3$. 
\textbf{c}, PT. 
\textbf{d}, BaTiO$_3$.
The x(y) direction refers to the in-plane direction.
The z direction refers to the out-of-plane direction.
}
\label{fig:lattice_1}
\end{figure}
\clearpage

\subsection{PtS$_2$ and Te$_2$I}
\begin{figure}[!htb]
\includegraphics[width=
\textwidth]{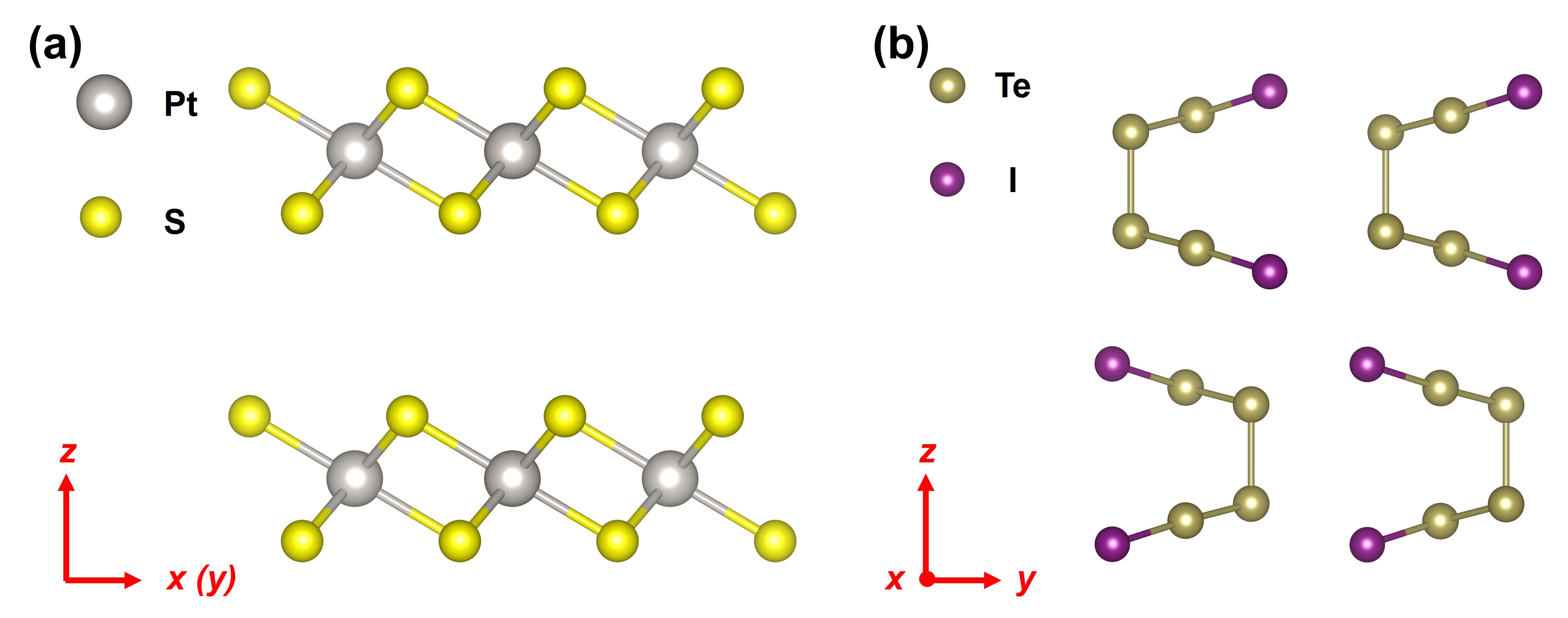}
\caption{Schematic illustrations of lattice structures and crystal orientations for the unit cells of PtS$_2$ and Te$_2$I. 
\textbf{a}, PtS$_2$. 
\textbf{b}, Te$_2$I. 
The z direction refers to the out-of-plane direction.
}
\label{fig:lattice_2}
\end{figure}
\clearpage

\section{VERIFICATION OF THERMODYNAMIC Relation FOR TYPICAL MATERIALS}
\subsection{Isotropic materials including CdS, CdTe, GaAs, and ZnSe}
\begin{figure}[!htb]
\includegraphics[width=
\textwidth]{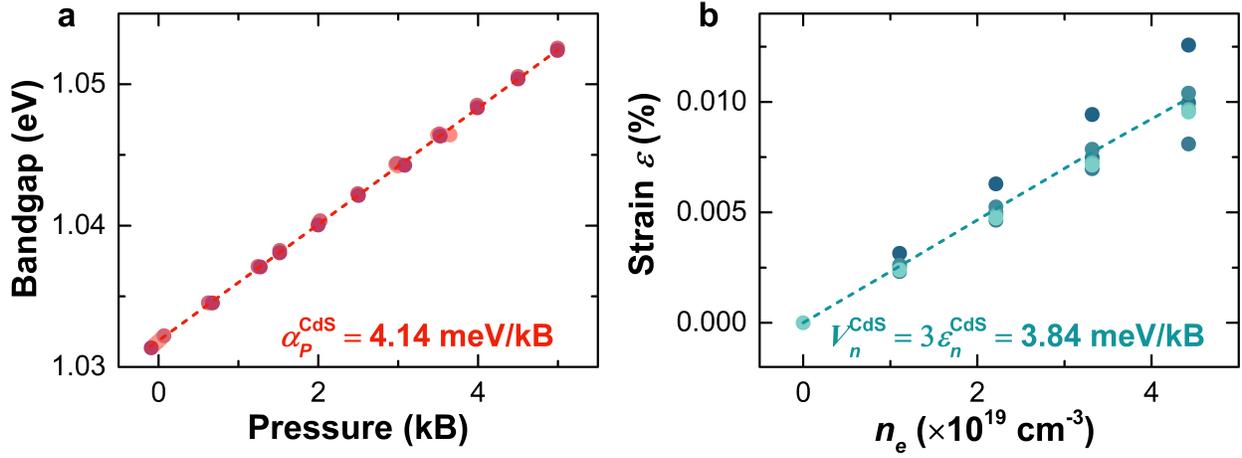}
\caption{\textbf{a}, CdS band gap change with hydrostatic pressure under different excited electron densities. 
Band gap pressure coefficient (dashed line) is fitted as 4.14 meV/kB.
\textbf{b}, CdS strain change with excited electron densities at different hydrostatic pressures.
Photostriction (dashed line) is fitted as 3.84 meV/kB.
}
\label{fig:CdS}
\end{figure}

\begin{figure}[!htb]
\includegraphics[width=
\textwidth]{Fig-SI/S_CdTe.png}
\caption{\textbf{a}, CdTe band gap change with hydrostatic pressure under different excited electron densities. 
Band gap pressure coefficient (dashed line) is fitted as 8.28 meV/kB.
\textbf{b}, CdTe strain change with excited electron densities at different hydrostatic pressures.
Photostriction (dashed line) is fitted as 8.17 meV/kB.
}
\label{fig:CdTe}
\end{figure}

\begin{figure}[!htb]
\includegraphics[width=
\textwidth]{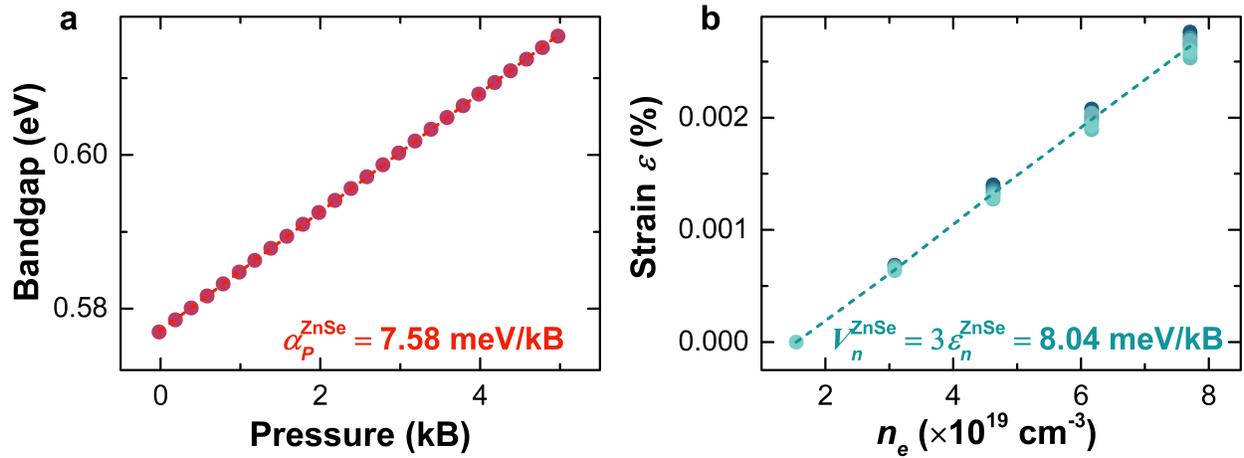}
\caption{\textbf{a}, ZnSe band gap change with hydrostatic pressure under different excited electron densities. 
Band gap pressure coefficient (dashed line) is fitted as 7.58 meV/kB.
\textbf{b}, ZnSe strain change with excited electron densities at different hydrostatic pressures.
Photostriction (dashed line) is fitted as 8.04 meV/kB.
}
\label{fig:ZnSe}
\end{figure}

\clearpage

\subsection{In-plane (r) directions of anisotropic material GaN}
\begin{figure}[!htb]
\includegraphics[width=
\textwidth]{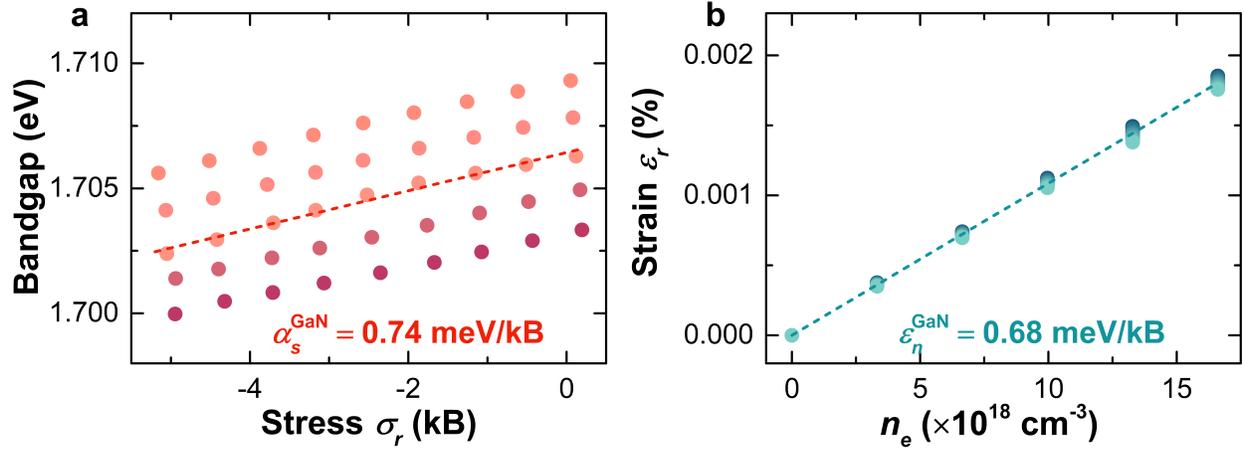}
\caption{\textbf{a}, GaN band gap change with stress along in-plane (r) direction (in-plane symmetry) under different excited electron densities. 
Band gap stress coefficient (dashed line) is fitted as 0.74 meV/kB.
\textbf{b}, GaN strain change along r direction with excited electron densities at different pressures.
Photostriction (dashed line) is fitted as 0.68 meV/kB.
}
\label{fig:MAPbI3_x}
\end{figure}
\clearpage

\subsection{In-plane (x and y) directions of anisotropic material MAPbI$_3$}
\begin{figure}[!htb]
\includegraphics[width=
\textwidth]{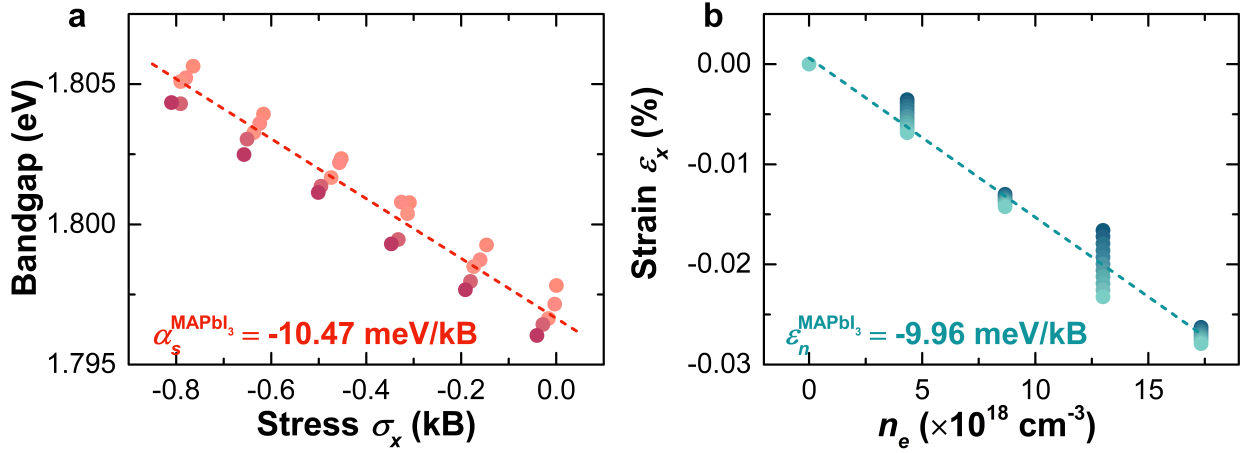}
\caption{\textbf{a}, MAPbI$_3$ band gap change with stress along in-plane (x) direction under different excited electron densities. 
Band gap stress coefficient (dashed line) is fitted as -10.47 meV/kB.
\textbf{b}, MAPbI$_3$ strain change along x direction with excited electron densities at different pressures.
Photostriction (dashed line) is fitted as -9.96 meV/kB.
}
\label{fig:MAPbI3_x}
\end{figure}

\begin{figure}[!htb]
\includegraphics[width=
\textwidth]{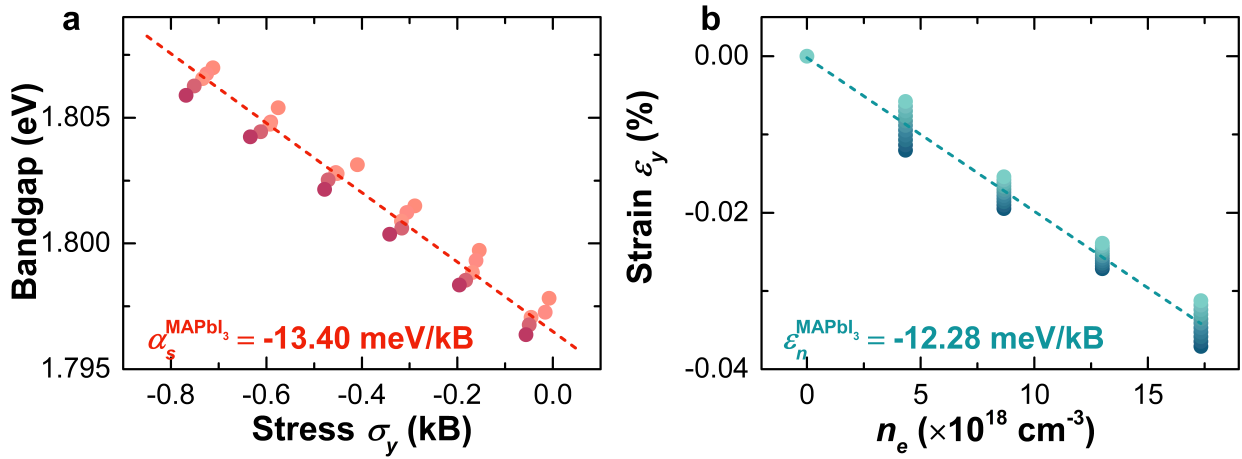}
\caption{\textbf{a}, MAPbI$_3$ band gap change with stress along in-plane (y) direction under different excited electron densities. 
Band gap stress coefficient (dashed line) is fitted as -13.40 meV/kB.
\textbf{b}, MAPbI$_3$ strain change along y direction with excited electron densities at different pressures.
Photostriction (dashed line) is fitted as -12.28 meV/kB.
}
\label{fig:MAPbI3_y}
\end{figure}
\clearpage

\section{ANALYSIS OF BAND GAP PRESSURE COEFFICIENTs FOR Discovered MATERIALS in high-throughput search}
\subsection{Relationship between band gap pressure coefficient and bulk modulus}
\begin{figure}[!htb]
\includegraphics[width=
\textwidth]{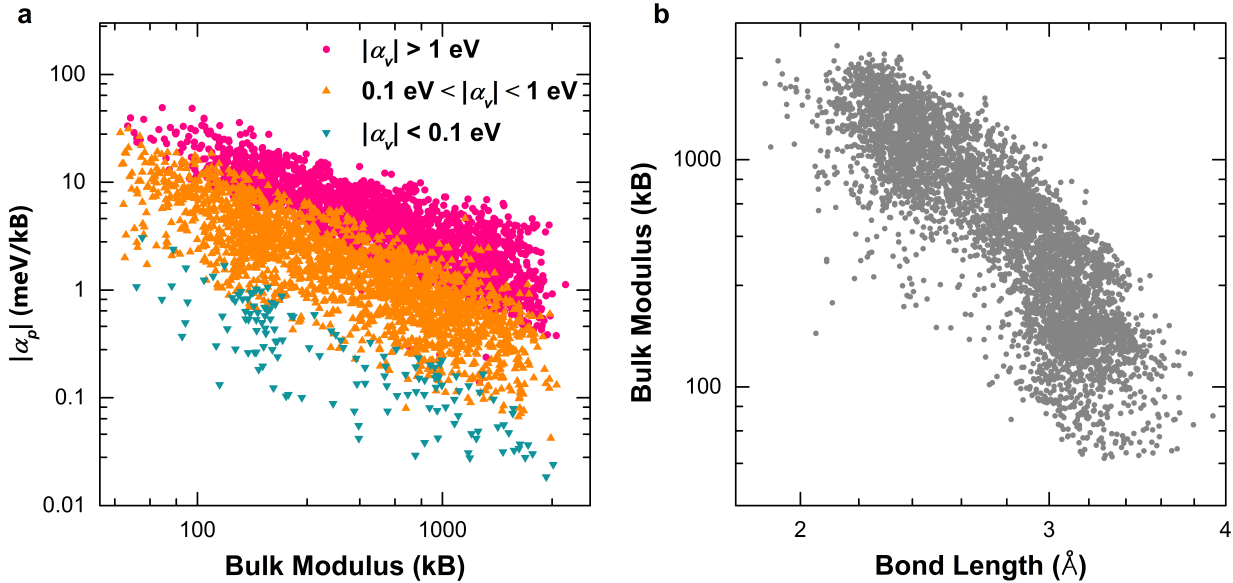}
\caption{a, Magnitude of band gap pressure coefficient with bulk modulus, segmented into three regions by the magnitude of $\alpha_v$. b, Change of bulk modulus with averaged bond length.}
\label{fig:pressure_coe}
\end{figure}

A decreasing trend in amplitude of band gap pressure coefficient with bulk modulus is observed after analyzing 4770 materials.
As presented in Fig. \ref{fig:pressure_coe}a, the decreasing trend becomes significant after segmenting the entire region based on the magnitude of $\alpha_V$.
Fig. \ref{fig:pressure_coe}b shows that the bulk modulus decreases with increasing bond length.
Taking this into consideration, it is reasonable to conclude photostriction (band gap pressure coefficient) generally shows an upward trend with increasing bond length.

\subsection{Candidate materials with large band gap pressure coefficient}
The following table contains the band gap pressure coefficient of top 500 stable materials with PBE-calculated band gap between 0 and 2 eV from the Materials Project\cite{jain2013commentary}. 
Band gap pressure coefficients are determined by linear regression of the band gap change with hydrostatic pressure (2 kB, 6 kB, and 10 kB).
The list is sorted by the amplitude of the band gap pressure coefficient.

\begin{longtable}{p{6cm}c}
         \caption{Top 500 photostrictive materials (with the Materials Project ID number).} \\
         \hline
         \ Material  & \thead{band gap Pressure Coefficient (meV/kB)}  \\
         \hline
         \endfirsthead
         \hline
         \ Material & \thead{band gap Pressure Coefficient (meV/kB)}  \\
         \hline
         \endhead
         \hline
         \endfoot
         PtS$_2$-mp$762$ & -49.2331 \\ 
         Te$_2$I-mp$27655$ & -48.5868 \\ 
         ScI$_3$-mp$1186991$ & -39.6388 \\ 
         HgI-mp$22859$ & -39.282 \\ 
         Te$_2$Br-mp$27648$ & -38.7564 \\ 
         AsS-mp$542810$ & -38.3489 \\ 
         PdSe$_2$-mp$2418$ & -35.824 \\ 
         WS$_2$-mp$224$ & -33.7557 \\ 
         PdS$_2$-mp$13682$ & -33.5788 \\ 
         PAuS$_4$-mp$30938$ & -33.2683 \\ 
         WSe$_2$-mp$1821$ & -33.0542 \\ 
         SiAs-mp$1863$ & -32.8593 \\ 
         Cs$_2$Se$_5$-mp$541055$ & -32.3042 \\ 
         AuSe-mp$2793$ & -31.6278 \\ 
         PI$_2$-mp$29443$ & -31.2949 \\ 
         InS-mp$19795$ & -31.1798 \\ 
         AsSe-mp$542570$ & -30.8151 \\ 
         MoSe$_2$-mp$1634$ & -30.1618 \\ 
         InSe-mp$20485$ & -29.8206 \\ 
         TlBr$_2$-mp$27398$ & -29.8043 \\ 
         CsSbS$_6$-mp$28701$ & -29.2265 \\ 
         Rb$_2$Se$_5$-mp$620372$ & -28.9617 \\ 
         Ga$_2$PdI$_8$-mp$30946$ & -28.9043 \\ 
         Se-mp$570481$ & -28.8784 \\ 
         Cs$_2$TlAsI$_6$-mp$1112544$ & -28.5609 \\ 
         SiAs$_2$-mp$978553$ & -28.4574 \\ 
         SnGeS$_3$-mp$5045$ & -28.0819 \\ 
         Cs$_2$S$_5$-mp$1193140$ & -28.0509 \\ 
         Rb$_2$TlSbCl$_6$-mp$1114002$ & -27.8811 \\ 
         CsK$_2$Sb-mp$581024$ & 27.6284 \\ 
         CdP$_1$$_4$Pb-mp$1214259$ & -27.1008 \\ 
         P$_3$Se$_4$I-mp$28393$ & -26.7091 \\ 
         Rb$_2$S$_5$-mp$16911$ & -26.6534 \\ 
         GeP-mp$1095275$ & -26.6284 \\ 
         P-mp$568348$ & -26.1563 \\ 
         Tl$_3$Cd$_2$I$_7$-mp$28432$ & -26.1179 \\ 
         SbI$_3$Cl$_8$-mp$23536$ & -25.7183 \\ 
         TiI$_4$-mp$541013$ & -25.615 \\ 
         NbI$_5$-mp$569578$ & -25.2136 \\ 
         CsGeI$_3$-mp$28377$ & -25.0838 \\ 
         SbI$_3$(BrCl$_3$)$_2$-mp$570962$ & -25.0286 \\ 
         Te$_2$Mo-mp$602$ & -24.7159 \\ 
         Pd(Se$_3$Br)$_2$-mp$2422143$ & -24.5998 \\ 
         CsK$_2$Bi-mp$863759$ & 23.9411 \\ 
         As$_2$Se$_3$-mp$909$ & -23.8915 \\ 
         InI$_2$-mp$29312$ & -23.6691 \\ 
         HI$_3$-mp$1184859$ & -23.4911 \\ 
         RbAu$_3$Se$_2$-mp$9385$ & -23.4292 \\ 
         CdSnP$_1$$_4$-mp$1199087$ & -23.2282 \\ 
         Nb$_2$Se$_1$$_7$Cl$_1$$_2$-mp$573340$ & -22.916 \\ 
         ThTe$_3$-mp$1025522$ & -22.7052 \\ 
         K$_2$Se$_5$-mp$18609$ & -22.5947 \\ 
         SiP$_2$-mp$9996$ & -22.0264 \\ 
         BP(IBr)$_3$-mp$567433$ & -21.9507 \\ 
         InBr-mp$22870$ & -21.8804 \\ 
         Rb$_3$Au$_3$Cl$_8$-mp$27301$ & -21.5006 \\ 
         K$_2$S$_5$-mp$17146$ & -21.4127 \\ 
         BiTeCl-mp$28944$ & -21.2176 \\ 
         CsTaI$_6$-mp$606510$ & -21.0869 \\ 
         CsInBr$_3$-mp$1112338$ & -20.9722 \\ 
         GeI$_4$-mp$23266$ & -20.9475 \\ 
         P$_3$BrCl$_1$$_4$-mp$29033$ & 20.702 \\ 
         Tl$_3$SbS$_4$-mp$8378$ & -20.6668 \\ 
         MoS$_2$-mp$1434$ & -20.6525 \\ 
         GeAs$_2$-mp$17524$ & -20.5343 \\ 
         K$_7$TaAs$_4$-mp$18073$ & 20.4017 \\ 
         Al$_2$Te$_3$-mp$29502$ & -20.3344 \\ 
         GaSe-mp$568263$ & -20.3016 \\ 
         Bi$_9$Ir$_2$I$_3$-mp$680182$ & -20.2553 \\ 
         KBaPSe$_4$-mp$18156$ & -20.0827 \\ 
         TlSb$_3$S$_5$-mp$27515$ & -19.5847 \\ 
         HfSnS$_3$-mp$8725$ & -19.3201 \\ 
         PICl$_6$-mp$27824$ & 19.1764 \\ 
         InI-mp$23202$ & -19.1215 \\ 
         TlI$_3$-mp$28329$ & -18.9936 \\ 
         HgI$_2$-mp$23192$ & -18.9549 \\ 
         ICl$_3$-mp$27729$ & -18.9476 \\ 
         C(Se$_2$Br)$_2$-mp$28131$ & -18.9361 \\ 
         PBr$_5$-mp$22874$ & 18.9155 \\ 
         Cs$_6$InAs$_3$-mp$22506$ & 18.8649 \\ 
         RbInCl$_3$-mp$1114001$ & -18.8596 \\ 
         RbAuBr$_3$-mp$27300$ & -18.8048 \\ 
         Tl$_5$Se$_2$Cl-mp$28920$ & -18.7913 \\ 
         TlCuPS$_3$-mp$1190539$ & -18.7649 \\ 
         TlSnPS$_4$-mp$6057$ & -18.7399 \\ 
         Hg$_3$TeBr$_4$-mp$29097$ & -18.7319 \\ 
         TlAuCl$_4$-mp$28368$ & -18.6798 \\ 
         PBr$_7$-mp$647343$ & 18.6185 \\ 
         Cs$_2$Te-mp$573763$ & 18.43 \\ 
         Tl$_3$BSe$_3$-mp$28490$ & -18.4141 \\ 
         Na$_3$AlAs$_2$-mp$8543$ & 18.3579 \\ 
         K$_2$NaAlP$_2$-mp$9068$ & 18.3427 \\ 
         Cs$_3$Na$_2$SnP$_3$-mp$17707$ & 18.2693 \\ 
         KNaTe-mp$8755$ & 18.2291 \\ 
         Tl$_2$Hg$_3$Se$_4$-mp$1105692$ & -18.1671 \\ 
         GaTeI$_7$-mp$29541$ & -18.1592 \\ 
         Bi$_2$Se$_3$-mp$541837$ & -18.1291 \\ 
         Rb$_2$P$_2$Pt-mp$1205641$ & -17.8567 \\ 
         Rb$_2$As$_2$Pt-mp$13445$ & -17.7568 \\ 
         K$_5$Tl$_3$O-mp$555531$ & 17.7068 \\ 
         InTeBr-mp$29236$ & -17.6732 \\ 
         AlHg$_2$SbCl$_4$-mp$568001$ & 17.6571 \\ 
         Cs$_2$SnBr$_6$-mp$641923$ & -17.6139 \\ 
         TlSe-mp$1836$ & -17.6069 \\ 
         K$_7$NbAs$_4$-mp$18361$ & 17.566 \\ 
         Cs$_3$BAs$_2$-mp$573658$ & 17.5432 \\ 
         Rb$_2$SnBr$_6$-mp$569028$ & -17.49 \\ 
         PPdSe-mp$3123$ & -17.4006 \\ 
         K$_2$NaAlAs$_2$-mp$9069$ & 17.3567 \\ 
         GePdS$_3$-mp$541785$ & -17.246 \\ 
         PdI$_2$-mp$27747$ & -17.0983 \\ 
         TaTeBr$_9$-mp$29716$ & -17.0979 \\ 
         Tl$_2$Sn$_2$S$_3$-mp$28011$ & -17.0852 \\ 
         Ga$_2$Te$_5$-mp$2371$ & -17.0675 \\ 
         GaTe-mp$542812$ & -16.9283 \\ 
         Na$_5$AuSe$_1$$_2$-mp$29198$ & -16.8263 \\ 
         K$_4$P$_2$$_1$I-mp$31280$ & -16.8147 \\ 
         CsSn-mp$571056$ & 16.7952 \\ 
         GeAs-mp$9548$ & -16.7525 \\ 
         GeSe-mp$700$ & -16.7325 \\ 
         BiI$_5$-mp$1214795$ & -16.6982 \\ 
         K$_2$As$_2$Pt-mp$1079513$ & -16.6839 \\ 
         Sn(BiTe$_2$)$_2$-mp$38605$ & -16.6398 \\ 
         Nb$_2$PS$_1$$_0$-mp$648932$ & -16.632 \\ 
         Rb$_2$Sb$_4$S$_7$-mp$561051$ & -16.6248 \\ 
         NaSbS$_2$-mp$5414$ & -16.6174 \\ 
         CsAuSe$_3$-mp$567913$ & -16.5906 \\ 
         NaSi-mp$2402$ & -16.5655 \\ 
         Hg(SbS$_2$)$_4$-mp$542596$ & -16.5612 \\ 
         TiP$_2$(S$_4$Cl)$_2$-mp$1203568$ & -16.542 \\ 
         Cs$_3$As(PSe$_3$)$_3$-mp$1201199$ & -16.5088 \\ 
         V$_2$Se$_9$-mp$28256$ & -16.5035 \\ 
         Tl$_2$PdCl$_4$-mp$29889$ & -16.384 \\ 
         GaHg$_2$SbCl$_4$-mp$571190$ & 16.3136 \\ 
         Rb$_5$GeP$_3$-mp$17978$ & 16.262 \\ 
         K$_2$AgBi-mp$27549$ & 16.2294 \\ 
         Tl$_4$PbSe$_3$-mp$1195087$ & -16.18 \\ 
         Na$_3$AlP$_2$-mp$5122$ & 16.1209 \\ 
         K$_2$AgAs-mp$7642$ & 16.1056 \\ 
         Rb$_3$BAs$_2$-mp$9718$ & 16.1027 \\ 
         K$_2$NaGaP$_2$-mp$9666$ & 15.9499 \\ 
         Si(PbSe$_2$)$_2$-mp$27532$ & -15.9444 \\ 
         CsHg$_2$Br$_5$-mp$619510$ & -15.8795 \\ 
         Tl$_5$Ag$_3$(PS$_3$)$_4$-mp$1195546$ & -15.8093 \\ 
         HgPS$_3$-mp$27178$ & -15.6762 \\ 
         K$_3$Au$_3$Br$_8$-mp$1211803$ & -15.6728 \\ 
         Ag$_2$TeS$_3$-mp$29163$ & -15.6409 \\ 
         HgP$_1$$_4$Pb-mp$22574$ & -15.6346 \\ 
         Ta$_2$Se$_1$$_7$Br$_1$$_2$-mp$541758$ & -15.6219 \\ 
         Rb$_2$PtI$_6$-mp$28110$ & -15.5877 \\ 
         AlTlSe$_2$-mp$863762$ & -15.5499 \\ 
         As$_2$S$_3$-mp$1189572$ & -15.519 \\ 
         TlAsSe$_2$-mp$30236$ & -15.4383 \\ 
         Tb$_2$Te$_3$-mp$1105684$ & -15.3691 \\ 
         InBi$_2$S$_4$Cl-mp$559521$ & -15.3461 \\ 
         SbAs$_2$Cl$_1$$_3$-mp$27786$ & 15.3279 \\ 
         Tl$_2$TeI$_6$-mp$31077$ & -15.3009 \\ 
         Cs$_2$Se-mp$1011695$ & 15.2859 \\ 
         CuTe$_2$Cl-mp$30971$ & -15.2785 \\ 
         K$_1$$_0$NbInAs$_6$-mp$1224638$ & 15.2746 \\ 
         ThSe$_3$-mp$1079673$ & -15.2388 \\ 
         AuCl$_2$-mp$541656$ & -15.2299 \\ 
         Tm$_2$Te$_3$-mp$1188133$ & -15.2129 \\ 
         AuCl$_3$-mp$27647$ & -15.0595 \\ 
         Er$_2$Te$_3$-mp$14643$ & -14.989 \\ 
         PdSCl-mp$560901$ & -14.9753 \\ 
         Rb$_3$BP$_2$-mp$9720$ & 14.9588 \\ 
         ICl-mp$567998$ & -14.9296 \\ 
         K$_3$Sb$_2$Au$_3$-mp$9273$ & -14.9144 \\ 
         Cs$_5$Tc$_6$S$_8$Br$_7$-mp$1182756$ & -14.9134 \\ 
         Rb$_2$SbAuS$_4$-mp$556894$ & -14.9117 \\ 
         KS$_3$-mp$1196514$ & -14.9102 \\ 
         Rb$_4$P$_2$$_1$I-mp$31279$ & -14.907 \\ 
         Tl$_3$AsS$_4$-mp$5704$ & -14.8805 \\ 
         Rb$_7$NbAs$_4$-mp$18440$ & 14.8642 \\ 
         Ho$_2$Te$_3$-mp$1189209$ & -14.8402 \\ 
         SnHg$_6$(As$_2$Br$_3$)$_2$-mp$567185$ & 14.8243 \\ 
         Cs$_2$AgAuCl$_6$-mp$567776$ & -14.8079 \\ 
         Tl$_3$AsSe$_3$-mp$7684$ & -14.784 \\ 
         Rb$_7$NaGe$_8$-mp$14407$ & 14.7421 \\ 
         Rb$_5$SiAs$_3$-mp$17490$ & 14.717 \\ 
         RbCaAs-mp$9845$ & 14.7149 \\ 
         Sn(S$_4$I)$_4$-mp$557101$ & -14.6968 \\ 
         RbGaH$_2$-mp$974682$ & 14.6531 \\ 
         Lu$_2$Te$_3$-mp$1189748$ & -14.6491 \\ 
         RbGe-mp$1888$ & 14.632 \\ 
         K$_3$As(PSe$_3$)$_3$-mp$1197620$ & -14.6292 \\ 
         Rb$_2$PbCl$_6$-mp$23475$ & -14.6234 \\ 
         Sn(SbTe$_2$)$_2$-mp$27947$ & -14.5844 \\ 
         ZrSnS$_3$-mp$17324$ & -14.5458 \\ 
         K$_2$NaGaAs$_2$-mp$9676$ & 14.5454 \\ 
         CsTlBr$_4$-mp$573702$ & -14.532 \\ 
         TlAgAs$_2$PbS$_5$-mp$677611$ & -14.4678 \\ 
         Cs$_2$PbCl$_6$-mp$23425$ & -14.4657 \\ 
         AlSiTe$_3$-mp$31220$ & -14.4188 \\ 
         RbSi-mp$1074$ & 14.4099 \\ 
         Cs$_2$NaGaP$_2$-mp$1213705$ & 14.4009 \\ 
         CsPSe$_6$-mp$1225967$ & -14.3896 \\ 
         Cs$_3$BrO-mp$1205638$ & 14.3249 \\ 
         K$_3$BP$_2$-mp$9664$ & 14.3117 \\ 
         Y$_2$Te$_3$-mp$1189791$ & -14.2932 \\ 
         Rb$_4$TlSbCl$_1$$_2$-mp$1219616$ & -14.2913 \\ 
         KSbS$_2$-mp$11703$ & -14.288 \\ 
         Tl$_2$GeS$_3$-mp$7277$ & -14.2671 \\ 
         InSbS$_3$-mp$21365$ & -14.2472 \\ 
         Rb$_7$NaSi$_8$-mp$1202504$ & 14.1719 \\ 
         Pr$_3$RuBr$_3$-mp$1209743$ & -14.1628 \\ 
         Dy$_2$Te$_3$-mp$1188427$ & -14.136 \\ 
         Nb$_2$Se$_9$-mp$541106$ & -14.1082 \\ 
         Cs$_2$Pt-mp$13548$ & 14.1007 \\ 
         MnV$_2$O$_6$-mp$2490654$ & -14.0129 \\ 
         Tb$_7$RhI$_1$$_2$-mp$1190279$ & -13.9526 \\ 
         In$_5$S$_5$Cl-mp$31263$ & -13.8979 \\ 
         K$_2$SnBr$_6$-mp$574040$ & -13.8762 \\ 
         Tl$_3$BS$_3$-mp$29337$ & -13.8151 \\ 
         K$_4$BeP$_2$-mp$9872$ & 13.7772 \\ 
         La$_3$RuI$_3$-mp$29824$ & -13.7687 \\ 
         Tl$_3$PSe$_4$-mp$4160$ & -13.7308 \\ 
         La(Tl$_3$Te$_2$)$_3$-mp$1223306$ & -13.7124 \\ 
         ZrHg$_3$(SBr$_3$)$_2$-mp$1207395$ & 13.7104 \\ 
         TlSbS$_2$-mp$676540$ & -13.6698 \\ 
         Rb$_4$SnSb$_6$-mp$505490$ & -13.6267 \\ 
         K$_3$InP$_2$-mp$20256$ & 13.6214 \\ 
         Li$_2$TeS$_3$-mp$558731$ & -13.6185 \\ 
         ErTl(PSe$_3$)$_2$-mp$1105213$ & -13.5431 \\ 
         RbAg$_2$(AsSe$_2$)$_3$-mp$570593$ & -13.5378 \\ 
         PtCl$_2$-mp$23290$ & -13.5248 \\ 
         KP$_2$Au$_5$S$_8$-mp$561218$ & -13.5236 \\ 
         Rb$_1$$_2$Sn$_2$As$_6$O-mp$1219774$ & 13.5164 \\ 
         PdCl$_2$-mp$29487$ & -13.5008 \\ 
         Rb$_2$SnAs$_2$-mp$8931$ & 13.4733 \\ 
         Te$_3$(PdBr)$_4$-mp$638749$ & -13.4665 \\ 
         TlGaTe$_2$-mp$3785$ & -13.458 \\ 
         K$_5$Au(IO)$_2$-mp$558332$ & 13.4438 \\ 
         InTeI-mp$29234$ & -13.4262 \\ 
         Tl$_5$Se$_2$Br-mp$28921$ & -13.3982 \\ 
         Tl$_4$SnS$_4$-mp$5818$ & -13.341 \\ 
         K$_3$Al$_2$As$_3$-mp$28347$ & 13.3196 \\ 
         K$_6$InAs$_3$-mp$1226185$ & 13.3089 \\ 
         Rb$_3$TlBr$_6$-mp$1114103$ & 13.2897 \\ 
         AlSb$_2$(TeCl$_2$)$_2$-mp$567743$ & -13.1588 \\ 
         Er$_7$IrI$_1$$_2$-mp$1105808$ & -13.1334 \\ 
         Re$_3$Br$_8$Cl-mp$1219236$ & -13.1128 \\ 
         K$_3$Na$_2$SnP$_3$-mp$16847$ & 13.1025 \\ 
         KPSe$_6$-mp$18625$ & -13.0885 \\ 
         CuP$_1$$_0$-mp$606644$ & -13.0816 \\ 
         Na$_5$InTe$_4$-mp$28597$ & -13.072 \\ 
         AgI-mp$580941$ & -13.0687 \\ 
         K$_4$ZnBi$_2$-mp$1103566$ & 13.0508 \\ 
         Rb$_3$As$_1$$_1$-mp$542233$ & -13.0106 \\ 
         RbAuSe-mp$9731$ & 12.9983 \\ 
         Rb$_3$ClO-mp$755428$ & 12.9871 \\ 
         MoBr$_3$O-mp$1221539$ & -12.9478 \\ 
         Tl$_6$SI$_4$-mp$27938$ & -12.921 \\ 
         Re$_3$Se$_2$Br$_5$-mp$1173441$ & -12.9061 \\ 
         Ba$_6$Sn$_6$Se$_1$$_3$-mp$1196224$ & -12.8915 \\ 
         Tl$_2$PtCl$_6$-mp$27834$ & -12.8757 \\ 
         PrTlTe$_2$-mp$999288$ & -12.874 \\ 
         Ga$_4$Hg$_1$$_1$(AsBr$_4$)$_4$-mp$669455$ & 12.8301 \\ 
         TlGeCl$_3$-mp$998744$ & -12.8083 \\ 
         EuI$_2$-mp$1102093$ & -12.7687 \\ 
         KAuCl$_4$-mp$27181$ & -12.7559 \\ 
         TlIn$_4$S$_5$Cl-mp$1208155$ & -12.7485 \\ 
         Nb$_2$Te$_6$I-mp$28745$ & -12.7322 \\ 
         K$_5$P$_2$Au-mp$14624$ & 12.7251 \\ 
         Ag$_2$HgI$_4$-mp$1229122$ & -12.7176 \\ 
         Cs$_2$As$_2$Pd-mp$8857$ & -12.6856 \\ 
         Rb$_5$FeO$_4$-mp$768830$ & 12.6704 \\ 
         NdTlTe$_2$-mp$999319$ & -12.6586 \\ 
         TeAuI-mp$27741$ & -12.6583 \\ 
         Y$_6$PtI$_1$$_0$-mp$1189258$ & -12.6408 \\ 
         K$_4$HgP$_2$-mp$8753$ & 12.6237 \\ 
         CuTe$_2$Br-mp$31036$ & -12.5825 \\ 
         K$_5$As$_2$Au-mp$8683$ & 12.5721 \\ 
         TiTlPS$_5$-mp$1542940$ & -12.5594 \\ 
         K$_2$BiAu-mp$1084770$ & 12.5432 \\ 
         ZrS$_2$-mp$1186$ & -12.528 \\ 
         Rb(GeAs)$_3$-mp$1192510$ & -12.5073 \\ 
         HfTl$_2$Se$_3$-mp$1102545$ & -12.4951 \\ 
         Rb$_2$S-mp$8041$ & 12.4915 \\ 
         Tl$_4$SiSe$_4$-mp$28334$ & -12.4784 \\ 
         K$_2$CuBr$_3$-mp$571440$ & 12.4649 \\ 
         Tb$_7$RhBr$_1$$_2$-mp$1189645$ & -12.4637 \\ 
         Na$_5$SiAs$_3$-mp$18139$ & 12.4617 \\ 
         Na$_6$GaP$_3$-mp$1173850$ & 12.4461 \\ 
         Rb$_2$Te$_5$-mp$31002$ & -12.4419 \\ 
         TaI$_5$-mp$570679$ & -12.439 \\ 
         Bi$_3$(TeCl$_5$)$_2$-mp$1193877$ & -12.4158 \\ 
         Tl$_2$PAuS$_4$-mp$9510$ & -12.3913 \\ 
         K$_3$Nb$_2$Se$_1$$_1$-mp$28428$ & -12.3893 \\ 
         Tl$_2$Ge$_2$Se$_5$-mp$540818$ & -12.3678 \\ 
         YTlTe$_2$-mp$1065514$ & -12.2823 \\ 
         K$_4$ZnP$_2$-mp$11719$ & 12.2259 \\ 
         TlPt$_2$S$_3$-mp$9272$ & -12.1989 \\ 
         SbSCl$_9$-mp$557809$ & -12.1871 \\ 
         Al(ICl$_2$)$_3$-mp$672352$ & -12.1854 \\ 
         PdBr$_2$-mp$27857$ & -12.1801 \\ 
         K$_2$Cu(PS$_3$)$_3$-mp$559644$ & -12.1655 \\ 
         Rb$_2$Te-mp$441$ & 12.1624 \\ 
         Pt(PbO$_2$)$_2$-mp$29332$ & -12.1479 \\ 
         ZrSe$_2$-mp$2076$ & -12.1434 \\ 
         TlSbAs$_2$PbS$_6$-mp$1208201$ & -12.1352 \\ 
         Rb$_4$SnO$_3$-mp$756570$ & 12.1291 \\ 
         CaPbI$_6$-mp$771691$ & -12.1217 \\ 
         Er$_7$CoI$_1$$_2$-mp$1105583$ & -12.1047 \\ 
         K$_4$HgAs$_2$-mp$29484$ & 12.1006 \\ 
         Rb$_6$Cl$_4$O-mp$29468$ & 12.0979 \\ 
         K$_7$NaGe$_8$-mp$1211990$ & 12.091 \\ 
         Ca$_2$Sb$_2$S$_5$-mp$29284$ & -12.0881 \\ 
         K$_4$Sn$_9$-mp$570900$ & -12.0836 \\ 
         RbHgSbTe$_3$-mp$1219595$ & -12.0753 \\ 
         DyTlTe$_2$-mp$1007908$ & -12.0478 \\ 
         K$_2$TeI$_6$-mp$27688$ & -12.0435 \\ 
         SnS-mp$2231$ & -12.0406 \\ 
         Na$_4$SnTe$_4$-mp$28108$ & -12.0396 \\ 
         Cs$_2$P$_2$Pd-mp$1205610$ & -12.005 \\ 
         WBr$_6$-mp$28483$ & -12.0044 \\ 
         Tl$_2$SiSe$_3$-mp$14241$ & -11.9981 \\ 
         TlRe$_3$(S$_2$Cl)$_2$-mp$560005$ & -11.9902 \\ 
         TlAsS$_2$-mp$4988$ & -11.9809 \\ 
         Nb$_3$I$_8$-mp$27772$ & 11.9197 \\ 
         Tb$_7$CoI$_1$$_2$-mp$1105706$ & -11.9134 \\ 
         K$_6$Na$_3$AlSb$_4$-mp$541707$ & 11.8905 \\ 
         Sr$_3$SbI$_3$-mp$1208807$ & -11.8664 \\ 
         GaAs-mp$2534$ & 11.8656 \\ 
         Sb$_8$(PbS$_5$)$_3$-mp$22737$ & -11.8652 \\ 
         TlSb(PSe$_3$)$_2$-mp$1105142$ & -11.8594 \\ 
         HgPSe$_3$-mp$7293$ & -11.8518 \\ 
         Cs$_2$PSe$_5$-mp$569060$ & -11.8509 \\ 
         Rb$_3$AgO$_2$-mp$553907$ & 11.8253 \\ 
         K$_4$ZnAs$_2$-mp$571461$ & 11.8187 \\ 
         Ge$_3$(Te$_3$As)$_2$-mp$541312$ & -11.8074 \\ 
         Rb$_3$LuCl$_6$-mp$1209449$ & 11.8036 \\ 
         Rb$_2$PdCl$_6$-mp$28145$ & -11.8002 \\ 
         Sn$_2$S$_3$-mp$1509$ & -11.7927 \\ 
         DyTl(PSe$_3$)$_2$-mp$1105239$ & -11.7899 \\ 
         Rb$_2$TeI$_6$-mp$28070$ & -11.7721 \\ 
         TlBi(PSe$_3$)$_2$-mp$567864$ & -11.6881 \\ 
         ErTlTe$_2$-mp$1007907$ & -11.6797 \\ 
         Er$_6$NiI$_1$$_0$-mp$1188388$ & -11.6791 \\ 
         Rb$_4$InSbCl$_1$$_2$-mp$1219602$ & -11.6694 \\ 
         KTlBr$_4$-mp$28048$ & -11.655 \\ 
         K$_2$NaInP$_2$-mp$21511$ & 11.6439 \\ 
         K$_3$Ga$_3$As$_4$-mp$567524$ & 11.6009 \\ 
         EuCl$_2$-mp$22887$ & -11.5921 \\ 
         SmTlTe$_2$-mp$999132$ & -11.5908 \\ 
         K$_2$NaInSb$_2$-mp$505767$ & 11.5757 \\ 
         Tl$_2$PSe$_3$-mp$28394$ & -11.5543 \\ 
         CsTaSe$_3$-mp$685607$ & -11.5286 \\ 
         CsSnSe$_3$-mp$1542038$ & -11.5186 \\ 
         AuCl-mp$32780$ & -11.5113 \\ 
         NiS$_2$-mp$1180046$ & -11.5039 \\ 
         Pd(SCl$_3$)$_2$-mp$28174$ & -11.4834 \\ 
         Sr$_3$P$_1$$_4$-mp$1853$ & -11.4379 \\ 
         GeBr$_2$-mp$541610$ & -11.4322 \\ 
         Ti$_6$Tl$_5$AgSe$_2$$_7$-mp$570021$ & -11.4312 \\ 
         K$_5$SnAs$_3$-mp$16989$ & 11.4099 \\ 
         K$_2$SeBr$_6$-mp$23036$ & -11.4059 \\ 
         Rb$_2$GaAs$_2$-mp$15416$ & 11.3956 \\ 
         AuBr-mp$505366$ & -11.3819 \\ 
         TmTlTe$_2$-mp$1008561$ & -11.3749 \\ 
         KSi-mp$1217$ & 11.3621 \\ 
         Rb$_6$Br$_4$O-mp$30005$ & 11.3592 \\ 
         Si(PbS$_2$)$_2$-mp$504564$ & -11.3555 \\ 
         ZrI$_3$-mp$570188$ & -11.3543 \\ 
         ZrP$_2$S$_7$-mp$31014$ & -11.3392 \\ 
         YTl(PSe$_3$)$_2$-mp$1105136$ & -11.3203 \\ 
         TaP$_2$Se$_2$Cl$_5$-mp$571250$ & -11.2988 \\ 
         K$_4$Ge(PSe$_3$)$_4$-mp$1195433$ & -11.2779 \\ 
         K$_2$PAuS$_4$-mp$9509$ & -11.2622 \\ 
         Ag$_2$GePbS$_4$-mp$861942$ & -11.2575 \\ 
         Cs$_7$NaSi$_8$-mp$1199908$ & 11.2299 \\ 
         TbTl(PSe$_3$)$_2$-mp$1105141$ & -11.2135 \\ 
         Rb$_2$Se-mp$11327$ & 11.1815 \\ 
         K$_2$NaInAs$_2$-mp$21510$ & 11.1629 \\ 
         RbTaGeS$_5$-mp$863022$ & -11.1581 \\ 
         Hg$_3$(SeBr)$_2$-mp$31075$ & -11.1528 \\ 
         Li$_1$$_0$Ge(PSe$_6$)$_2$-mp$721239$ & -11.1486 \\ 
         Li$_1$$_0$Si(PSe$_6$)$_2$-mp$721253$ & -11.1318 \\ 
         Na$_8$TiAs$_4$-mp$9071$ & 11.1292 \\ 
         In$_4$SnS$_4$-mp$646878$ & 11.1081 \\ 
         Cs$_2$PtBr$_6$-mp$30062$ & -11.1001 \\ 
         HfSe$_2$-mp$985831$ & -11.0668 \\ 
         In$_5$S$_5$Br-mp$510347$ & -11.0602 \\ 
         Tl$_3$AsSe$_4$-mp$1201566$ & -11.0465 \\ 
         CuP$_4$S$_3$I-mp$554724$ & -11.034 \\ 
         Rb$_2$Cr(BrCl)$_2$-mp$1219755$ & -11.0243 \\ 
         PtI$_4$-mp$669496$ & -11.0132 \\ 
         BaAg$_2$SnSe$_4$-mp$569114$ & -10.9928 \\ 
         AlAg(PSe$_3$)$_2$-mp$1228964$ & -10.9902 \\ 
         K$_2$Ga$_2$As$_3$-mp$15432$ & 10.9594 \\ 
         TlIn$_5$S$_6$-mp$505747$ & 10.9049 \\ 
         Rb$_3$Sb$_2$I$_9$-mp$1209693$ & -10.8953 \\ 
         K$_2$NaAs-mp$34127$ & 10.8794 \\ 
         Rb$_2$PdBr$_6$-mp$28084$ & -10.8576 \\ 
         TlInTe$_2$-mp$22791$ & -10.8527 \\ 
         Tl$_2$Te$_3$-mp$29711$ & -10.8432 \\ 
         Tl$_2$GeSe$_3$-mp$14242$ & -10.785 \\ 
         CrAg(PSe$_3$)$_2$-mp$1226281$ & -10.7691 \\ 
         Cs$_2$NaAlH$_6$-mp$989642$ & 10.7451 \\ 
         RbNb$_4$Ag(PS$_1$$_0$)$_2$-mp$1219639$ & -10.74 \\ 
         NbTeI$_3$-mp$540924$ & -10.7365 \\ 
         SiAg$_6$HgSe$_6$-mp$1219355$ & 10.7348 \\ 
         Cs(SbSe$_2$)$_2$-mp$3312$ & -10.723 \\ 
         Na$_3$Sb-mp$7956$ & 10.6796 \\ 
         SbAsPCl$_1$$_3$-mp$569209$ & -10.6534 \\ 
         Tl$_3$AgTe$_2$-mp$650442$ & -10.6447 \\ 
         Tl$_3$GeTe$_3$-mp$17217$ & -10.6423 \\ 
         TeRhCl-mp$22945$ & -10.6422 \\ 
         MoCl$_6$-mp$1221436$ & -10.641 \\ 
         Tl$_3$SbS$_3$-mp$8393$ & -10.6005 \\ 
         Rb$_3$LiPb$_4$-mp$1209665$ & 10.5982 \\ 
         Rb$_5$Ti$_6$AgSe$_2$$_7$-mp$16001$ & -10.5834 \\ 
         WCl$_4$-mp$980949$ & -10.5642 \\ 
         NaCaAs-mp$961685$ & 10.5456 \\ 
         K$_1$$_1$Cd$_2$Sb$_5$-mp$1194660$ & 10.5373 \\ 
         LuTlTe$_2$-mp$1017508$ & -10.526 \\ 
         Rb$_3$Tm$_2$Cu$_4$Br$_1$$_3$-mp$623862$ & 10.518 \\ 
         CsVI$_3$-mp$22991$ & -10.5036 \\ 
         Y$_6$NiI$_1$$_0$-mp$1188535$ & -10.5035 \\ 
         K$_2$GaAs$_2$-mp$15415$ & 10.4982 \\ 
         K$_2$PtCl$_6$-mp$23513$ & -10.4922 \\ 
         ZrHg$_3$(SeBr$_3$)$_2$-mp$569970$ & 10.4879 \\ 
         HgBr$_2$-mp$571558$ & -10.4851 \\ 
         ZrSe$_3$-mp$1683$ & -10.4673 \\ 
         KAuSe-mp$9881$ & 10.419 \\ 
         Cs$_2$PdBr$_6$-mp$1205847$ & -10.4175 \\ 
         K$_2$AsAuS$_4$-mp$9511$ & -10.3994 \\ 
         Nb$_4$P$_2$S$_2$$_1$-mp$1220713$ & -10.3922 \\ 
         KTa(AgSe$_2$)$_2$-mp$571288$ & -10.3775 \\ 
         LiAuS-mp$29829$ & -10.3655 \\ 
         Rb$_2$Hg$_3$Te$_4$-mp$29107$ & -10.3603 \\ 
         Na$_1$$_0$NbGaAs$_6$-mp$1221463$ & 10.3427 \\ 
         Al$_2$Te$_5$-mp$9254$ & -10.3309 \\ 
         ZnP$_1$$_4$Pb-mp$1197426$ & -10.3224 \\ 
         TlGaSe$_2$-mp$17254$ & -10.3172 \\ 
         NdTlSe$_2$-mp$568588$ & -10.3171 \\ 
         Ba$_2$P$_7$Br-mp$1105535$ & -10.3056 \\ 
         PrTl(PSe$_3$)$_2$-mp$1195020$ & -10.2884 \\ 
         K$_2$WSe$_4$-mp$18138$ & -10.2666 \\ 
         Tl$_2$S$_5$-mp$30520$ & -10.2536 \\ 
         BSe$_2$-mp$1214789$ & -10.2397 \\ 
         K$_3$Sn$_2$P$_5$S$_1$$_8$-mp$1211845$ & -10.2149 \\ 
         Ga$_2$PdBr$_8$-mp$30945$ & -10.2131 \\ 
         Cs$_4$FeO$_3$-mp$1205381$ & 10.2058 \\ 
         Na$_6$GaAs$_3$-mp$1173830$ & 10.2055 \\ 
         In$_4$Te$_3$-mp$617281$ & 10.2046 \\ 
         Ba(SbS$_2$)$_2$-mp$28129$ & -10.1944 \\ 
         AuSCl$_7$-mp$556587$ & -10.1809 \\ 
         LiAuI$_4$-mp$29520$ & -10.1766 \\ 
         K$_3$In$_2$As$_3$-mp$583615$ & 10.169 \\ 
         Ag$_3$AsSe$_3$-mp$662599$ & 10.1514 \\ 
         K$_2$Y$_4$Sn$_2$S$_1$$_1$-mp$560785$ & -10.1365 \\ 
         SnSe-mp$691$ & -10.1275 \\ 
         NaSrAs-mp$9775$ & 10.1193 \\ 
         Tl$_3$SiTe$_3$-mp$568020$ & -10.0861 \\ 
         Ba$_4$Sb$_4$Se$_1$$_1$-mp$28238$ & -10.0826 \\ 
         K$_5$CuAs$_2$-mp$14623$ & 10.0685 \\ 
         PrTlSe$_2$-mp$999289$ & -10.067 \\ 
         K$_2$Sn(AsS$_3$)$_2$-mp$10776$ & 10.0639 \\ 
         Pr$_3$S$_3$N-mp$558940$ & -10.0506 \\ 
         AgAuCl$_4$-mp$570340$ & -10.0434 \\ 
         KNb$_2$PS$_1$$_0$-mp$542972$ & -10.0408 \\ 
         K$_2$Pt-mp$1062676$ & 10.0286 \\ 
         Pr$_3$OsI$_3$-mp$570779$ & -10.0203 \\ 
         ZrTl$_2$S$_3$-mp$1102830$ & -10.0137 \\ 
         Tl$_9$SbSe$_6$-mp$676274$ & 10.0131 \\ 
         BaNaBi-mp$31235$ & 9.9952 \\ 
         Tl$_3$SbSe$_3$-mp$4876$ & -9.9764 \\ 
         K$_3$IO-mp$28171$ & 9.9758 \\ 
         K$_2$PtBr$_6$-mp$27691$ & -9.9723 \\ 
         Li$_2$WS$_4$-mp$753195$ & -9.9552 \\ 
         Rb$_3$P$_1$$_1$-mp$1197062$ & -9.9397 \\ 
         Ga$_2$TeSe$_2$-mp$28423$ & 9.9353 \\ 
         TeSe$_2$-mp$1217269$ & -9.9322 \\ 
         ZnTe-mp$2176$ & 9.9257 \\ 
         LuNiBi-mp$30457$ & 9.9053 \\ 
         In$_2$Te$_5$-mp$1197742$ & -9.8861 \\ 
         NbTlBr$_4$O-mp$551826$ & -9.8804 \\ 
         SbICl$_8$-mp$569512$ & -9.8747 \\ 
         K$_3$CuO$_2$-mp$28273$ & 9.8416 \\ 
         K$_2$Te$_2$As-mp$29380$ & -9.8289 \\ 
         K$_2$ZrTe$_3$-mp$29602$ & -9.8276 \\ 
         K$_2$PdCl$_6$-mp$23067$ & -9.7821 \\ 
         Rb$_3$H$_1$$_2$N$_5$-mp$1198991$ & 9.7717 \\ 
         CsAlTe$_3$-mp$1189272$ & -9.7394 \\ 
         K$_2$TeSe$_3$-mp$28419$ & -9.7342 \\ 
         NaInSn$_2$-mp$1220784$ & 9.732 \\ 
         K$_6$InP$_3$-mp$1224505$ & 9.711 \\ 
         CdPSe$_3$-mp$1079559$ & 9.7104 \\ 
         SmTlSe$_2$-mp$999137$ & -9.6776 \\ 
         Al(TeCl$_2$)$_2$-mp$540628$ & -9.661 \\ 
         KInSnSe$_4$-mp$568379$ & -9.6565 \\ 
         YAsSe-mp$1095603$ & -9.644 \\ 
         Ba(SbSe$_2$)$_2$-mp$4727$ & -9.636 \\ 
         LiHoSe$_2$-mp$15796$ & -9.631 \\ 
         RbAuI$_3$-mp$568666$ & -9.623 \\ 
         Sr(InP)$_2$-mp$1078973$ & 9.6046 \\ 
         KMgP-mp$1018737$ & 9.594 \\ 
         Rb$_2$MoSe$_4$-mp$1193232$ & -9.5881 \\ 
         \hline
         
\end{longtable}

\bibliography{references.bib}